\documentclass[lettersize,journal]{IEEEtran}
\IEEEoverridecommandlockouts
\usepackage{amsmath,amsfonts,amsthm,amssymb}
\usepackage{array}
\usepackage{textcomp}
\usepackage{subfigure}
\usepackage{float}
\usepackage{enumerate}
\usepackage{tabularx}
\usepackage{stfloats}
\usepackage{url}
\usepackage{hyperref}
\usepackage{xcolor,soul,framed} 
\usepackage{pstricks}
\usepackage{verbatim}
\usepackage{graphicx}
\usepackage{epstopdf}
\usepackage{booktabs}
\usepackage{cite}
\usepackage[T1]{fontenc}
\usepackage{tikz}
\usepackage{algorithm}
\usepackage{algpseudocode}
\usepackage{setspace}
\usepackage{booktabs}
\usepackage{makecell}
\usepackage{multirow}
\allowdisplaybreaks
\begin{document}
\newtheorem{Remark}{Remark}
\newtheorem{remark}{Remark}
\newtheorem{thm}{Theorem}
\renewcommand{\algorithmicrequire}{\textbf{Input:}} 
\renewcommand{\algorithmicensure}{\textbf{Output:}}

\title{Fair Beam Allocations through Reconfigurable Intelligent Surfaces\\
\thanks{This work is supported by the National Natural Science Foundation of China (NSFC) under Grants NO.12141107 and NO.11801200, and the Interdisciplinary Research Program of HUST (2023JCYJ012).}
}

%
\author{Rujing~Xiong,
        Ke Yin, Tiebin Mi,~Jialong~Lu,~Kai Wan,
        Robert~Caiming~Qiu,~\IEEEmembership{Fellow,~IEEE,}
\thanks{R.~Xiong, T. Mi, J. Lu, and R. Qiu are with the School of Electronic Information and Communications, Huazhong University of Science and Technology, Wuhan 430074, China (e-mail: \{rujing,mitiebin,m202272434,caiming\}@hust.edu.cn).}
\thanks{K. Yin is with the Center for Mathematical Sciences, Huazhong University of Science and Technology, Wuhan 430074, China (e-mail: kyin@hust.edu.cn).}
\thanks{K. Wan is with the School of Electronic Information and Communications, Huazhong University of Science and Technology, Wuhan 430074, China (e-mail: istwankai2@gmail.com).}
\thanks{A short version of this manuscript has been submitted to IEEE ICC 2024.}

}
\maketitle



\maketitle

\begin{abstract}
A fair beam allocation framework through reconfigurable intelligent surfaces (RISs) is proposed, incorporating the Max-min criterion. This framework focuses on designing explicit beamforming functionalities through optimization. Firstly, realistic models, grounded in geometrical optics, are introduced to characterize the input/output behaviors of RISs, effectively bridging the gap between the requirements on explicit beamforming operations and their practical implementations. Then, a highly efficient algorithm is developed for Max-min optimizations involving quadratic forms. Leveraging the Moreau-Yosida approximation, we successfully reformulate the original problem and propose iterations to attain the optimal solution. A comprehensive analysis of the algorithm's convergence is provided. Importantly, this approach exhibits excellent extensibility, making it readily applicable to address a broader class of Max-min optimization problems. Finally, numerical and prototype experiments are conducted to validate the effectiveness of the framework. With the proposed beam allocation framework and algorithm, we clarify that several crucial redistribution functionalities of RISs, such as explicit beam-splitting, fair beam allocation, and wide-beam generation, can be effectively implemented. These explicit beamforming functionalities have not been thoroughly examined previously. 

\end{abstract}


\section{Introduction}
\IEEEPARstart{R}{econfigurable} intelligent surfaces (RISs) have recently demonstrated profound potential in reconfiguring wireless propagation environments~\cite{wu2019towards,basar2019wireless,di2020smart}. What distinguishes this technique from other competitive technologies is its capability to provide passive relays through flexible beamforming functions. RISs consist of cost-effective, well-designed electromagnetic units, each capable of independently manipulating the characteristics of incident electromagnetic (EM) waves, including phase, amplitude, and polarization. Such configurations enable the redistribution of incident waves, allowing for the realization of flexible reflective radiation patterns, a key feature of RIS applications.

Extensive investigations have been conducted on the performance of RISs across various wireless systems, highlighting their effectiveness in redistributing EM waves. These systems encompass diverse scenarios, including but not limited multi-antenna and/or multi-user communication~\cite{di2020hybrid,chen2023channel}, physical-layer security~\cite{yang2020secrecy,cui2019secure}, orthogonal frequency division multiplexing-based communication~\cite{lin2020adaptive,yang2020intelligent}, unmanned aerial vehicle communication and networks~\cite{li2020reconfigurable,mu2021intelligent}, simultaneous wireless information and power transfer systems~\cite{pan2020intelligent,wu2020joint}, and mobile edge computing~\cite{bai2020latency,mao2022reconfigurable}, among others.

From a mathematical perspective, exploring beamforming through RISs requires investigation into two critical aspects. The first involves the scope of wave redistribution achievable via RISs without delving into specific algorithms.  
The fundamental problems are whether RISs can effectively redistribute arbitrary incident EM waves to achieve any desired scattered pattern and, the potential range of reflective patterns. 
Our previous study highlighted that, at least for uniform linear RISs, achieving arbitrary redistribution is not feasible due to limitations in the phase configurations \cite{mi2023towards}. The scope of beamforming through RISs is often overlooked and has not been conclusively clarified.  This underscores a critical challenge in the practical deployment of RISs.


The second aspect involves developing algorithms for optimal RIS phase configurations to meet specific beamforming requirements. These algorithms can be categorized into two types in general. 
On one hand, the phase compensation approach~\cite{tang2020wireless,pei2021ris,liang2022backscatter,xiong2023ris} stands out as one of the most straightforward techniques due to its low computational complexity and high physical interpretability. This approach is grounded in the observation that, for narrowband signals, the time differences resulting from multiple scattered paths can be effectively compensated by phase shifting. A fundamental limitation of this technique is that continuous phase compensation configurations encounter a substantial constraint known as the anomalous mirror effect, especially in the presence of multiple incident EM waves~\cite{mi2023towards}. 

Additionally, optimization-based approaches have been employed to achieve complex beamforming functions. In scenarios with multiple users, two strategies are frequently employed: the sum rate maximizing criterion \cite{huang2018achievable, peng2022performance, dong2023robust, shen2023multi} or Max-min criterion to ensure fairness \cite{kammoun2020asymptotic, xie2020max}. The max-min fairness approach involves solving semi-infinite Max-min problems. One existing approach is to transform the Max-min problems into constrained optimization problems and solve them using nonlinear programming techniques\cite{cai2011unified}. To reduce the computation cost, another approach treats it as a minimization problem of a maximum-type function. Since the maximum function is non-smooth, they use a projected sub-gradient algorithm to solve it \cite{xie2020max}. A third approach turns the minimax problem into the goal attainment problem, solved by the sequentially quadratic programming (SQP) method \cite{brayton1979new}. Besides classical optimization approaches, there are also heuristic algorithms that combine the greedy algorithm with randomized coordinate descent \cite{subhash2023max}.  

The phase compensation approach, grounded in a physical model, is generally constrained to single-input-single-output (SISO) scenarios. For scenarios involving multiple-input-multiple-output (MIMO), the optimization-based approach typically demands precise estimation of channel state information (CSI) due to the statistical channel assumptions, which is particularly challenging due to the absence of dedicated signal processing capabilities in passive RIS units. In RIS-aided MIMO communications, effectively harnessing physical modeling for RIS phase optimization to bypass intricate channel estimations remains a challenging and unresolved task.

\subsection{Contributions}
For explicit beamforming functionalities, we refer to beamforming independent of channel estimations. This study initiates an exploration of the explicit redistributor function of RISs, with a specific emphasis on fair beam allocations for multiuser through a single RIS. The main contributions are summarized as follows:
\begin{itemize}
\item \textbf{Geometrical optics based model and problem formulation}. 
In contrast to the prevalent use of statistical approaches in modeling the interactions between RISs and EM waves, this study presents more practical and realistic models grounded in geometrical optics to characterize the input/output behaviors of RISs. The adoption of these models aims to bridge the gap between theoretical analysis and practical beamforming implementations. We further tailor the general model to diverse scenarios, resulting in more concise models and enhancing the applicability of the proposed approach. To explore RISs' beamforming functionality, a fair beam allocation problem based on Max-min optimizations is formulated.

\item \textbf{Moreau-Yosida approximation-based (MA) method for Max-min problems}. 
Highly efficient algorithms are developed for Max-min optimizations involving quadratic forms. In this context, we leverage the Moreau-Yosida approximation~\cite{yosida2012functional} to reformulate the original optimization problem and propose iterations to attain the optimal solution. Additionally, we provide a comprehensive analysis of the algorithm's convergence. We prove that the proposed algorithm converges to a $\epsilon$-approximate first order optimal point within $O(\log(1/\epsilon))$ iterations. In each iteration, a smooth unconstrained sub-problem is solved using an accelerated gradient descent algorithm. This approach is intricately connected to the Majorization Minimization (MM) algorithmic framework and exhibits excellent extensibility, making it readily applicable to address a broader class of Max-min optimization problems.

\item \textbf{Flexible beam allocation functionalities exploration}. 

We investigate the beam redistribution functions of RISs using the proposed Moreau-Yosida approximation-based (MA) method. These explicit beamforming functionalities have not been thoroughly examined previously. In particular, we clarify that essential functions such as beam-splitting, fair beam allocation, and wide-beam generation can be effectively implemented within this framework. 
Additionally, the fair beam allocation framework is validated through prototype experiments.
\end{itemize}
%
%
%
%
%
%
%
%
%
%
%
%


\subsection{Outline}
The remainder of the paper is organized as follows. In Section~\ref{Section2}, we present the modeling of RIS-aided multi-user communications and formulate the beamforming problems for maximizing the minimum received signal power among users. Section~\ref{Section3} proposes an efficient algorithm along with Moreau-Yosida approximation to solve the problem. In Section~\ref{Section4}, We validate various beam allocation functions of RISs, such as beam splitting, fair beam allocation, and wide-beam generation. Section~\ref{Section5} is dedicated to the performance evaluations of the proposed algorithm through numerical and experimental tests. Finally, the paper is concluded in Section~\ref{Section6}.


\subsection{Notations}
The imaginary unit is indicated by $j$. The magnitude, real and complex components of a complex number are represented by $|\cdot|$, $\mathcal{\Re}(\cdot)$ and $\mathcal{\Im}(\cdot)$, respectively. $D_x$ represents the gradient concerning $x$. Unless explicitly specified, lower and upper case bold letters denote vectors and matrices. The conjugate transpose, conjugate, and transpose of~$\mathbf{A}$ are represented as $\mathbf{A}^H$, $\mathbf{A}^{*}$ and $\mathbf{A}^T$, respectively. 

\section{Physical modeling and problem formulation}\label{Section2}

\begin{figure}[!htbp]
  \centering
  \includegraphics[width=0.8\linewidth]{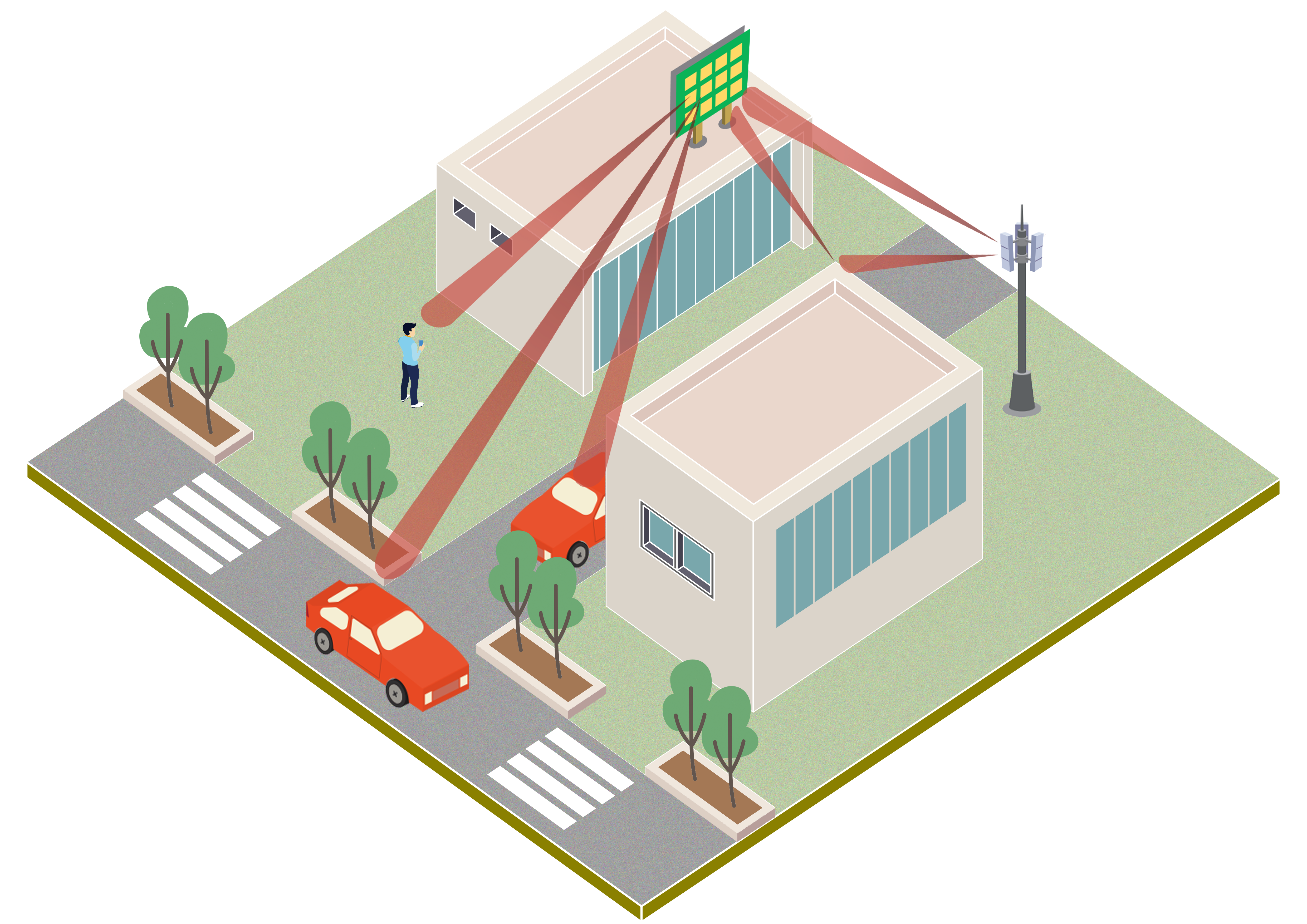}
  \caption{RIS-aided multi-user wireless communication.}
  \label{Scenarios}
\end{figure}

In the multi-user downlink communications scenario depicted in Fig.~\ref{Scenarios}, A reconfigurable intelligent surface (RIS) consisting of $N$ reflective units is employed to improve the communication quality of users. The RIS interacts with multipath signals from the transmit antennas or the strong reflections by other objects. For clarify and simplify, these potent multipath EM signals are regarded as $M$ independent incident source (or transmit) signals throughout this paper.

Two primary methodologies exist for modeling the interactions between RISs and EM waves~\cite{xiong2023design}. The first employs a statistical approach to directly estimate the channel state information (CSI). A significant observation is that the RIS-assisted channel functions as a cascade of two subchannels. The large-scale propagation path is influenced by the EM properties of the RIS, leading to substantial deviations from conventional statistical assumptions. Meanwhile, this complexity amplifies the challenge in the estimation of statistical CSI, primarily due to the absence of dedicated signal processing capabilities in passive RIS units.

The second approach is rooted in physics and draws direct inspiration from EM theory. Specifically within the context of RIS-aided systems, the response of the RIS can be calculated as the product of the isolated unit's scattering pattern and the response of the isotropic reflective array under mild assumptions. In the concept of an isotropic reflective array, the units are hypothetically replaced with isotropic scattering elements. This multiplication law for patterns is validated in traditional antenna arrays \cite{stutzman2012antenna}.


To characterize the input/output behaviors of the entire array, we utilize geometrical optics in~\cite{mi2023towards}. The reason behind employing geometrical optics is its precision when the size of the object is significantly larger than the wavelength, a condition met by RISs consisting of hundreds to thousands of units~\cite{mi2023towards}. For intuitive understanding, phase discrepancies arising from configurations and interelement path length differences play a pivotal role in beamforming functionality. These path length differences are closely related to the topology and geometry of RISs. 
Geometrical optics offers a high-precision method for calculating the phase discrepancy resulting from interelement path length differences and effectively bridges the gap between the requirements for explicit beamforming operations and their practical implementations.

\subsection{The Response of RISs to External Illuminations}

\begin{figure}[!htbp]
  \centering
  \includegraphics[width=0.9\linewidth]{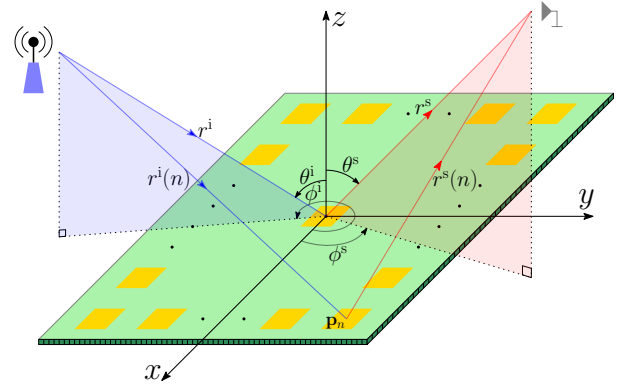}
  \caption{The RIS is illuminated by a single incident EM wave from $(r^\text{i}, \theta^\text{i}, \phi^\text{i})$.}
  \label{F:ArrayGeometry}
\end{figure}

For our analysis, we first concentrate on an arbitrary RIS situated on the $xoy$-plane, comprising multiple units located at $\mathbf{p}_n = [x_n, \ y_n, \ z_n ]^T$, $n=1, \ldots, N$. The RIS is illuminated by a single incident EM wave originating from $(r^\text{i}, \theta^\text{i}, \phi^\text{i})$. For clarity, $r^\text{i}$ denotes the distance from the source to the anchor of the RIS, such as the geometrical center for a planar array or the left end for a uniform linear array, while $r^\text{i} (n)$ represents the distance from the source to the $n$-th unit at $\mathbf{p}_n$, $ (\theta^\text{i}, \phi^\text{i})$ denotes the elevation and azimuth angles, as illustrated in Fig.~\ref{F:ArrayGeometry}. 

In the case of a point source, radiation propagates radially. As the EM wave propagates toward the unit at $\mathbf{p}_n$, the attenuation behavior is characterized by the factor $r^\text{i} (n) e^{ j 2 \pi r^\text{i} (n) / \lambda }$. The incident electric field at $\mathbf{p}_n$, denoted by $E^\text{i} (n) $, could be represented as $E^\text{i} ( r^\text{i}, \theta^\text{i}, \phi^\text{i} ) e^{ - j 2 \pi r^\text{i} (n) / \lambda}/ r^\text{i} (n)$. Similarly, if $E^\text{s} (n)$ denotes the scattered electric field of the unit at $\mathbf{p}_n$, then the electric field at the observation point is $E^\text{s} (n) e^{ - j 2 \pi r^\text{s} (n) / \lambda } / r^\text{s} (n)$. 

We now examine the scattering pattern of an isolated unit denoted as $\tau(\theta^\text{s}, \phi^\text{s}; \theta^\text{i}, \phi^\text{i})$. This notation signifies the dependence on both incident angle $(\theta^\text{i}, \phi^\text{i})$ and scattered angle $(\theta^\text{s}, \phi^\text{s})$. Utilizing the physical optics method, we calculate the bistatic scattered field of a rectangular metallic patch with near-zero thickness, as elaborated in \cite{mi2023towards}. It's worth noting that, for some units, a main resonant metallic rectangular patch does not exist. Analytically calculating this function presents a considerable challenge.

Combining the three components along the propagation trace through the $n$-th unit at $\mathbf{p}_n$, the electric field at the observation point is expressed  as 
\begin{multline}\label{SISOBehavior}
  E^\text{s}_n ( r^\text{s}_n, \theta^\text{s}, \phi^\text{s} ) = \tau ( \theta^\text{s} (n), \phi^\text{s} (n) ; \theta^\text{i} (n), \phi^\text{i} (n) ) e^{j \omega_n} \\
  \frac{ e^{ - j 2 \pi r^\text{i} (n) / \lambda} e^{ - j 2 \pi r^\text{s} (n) / \lambda } }{ r^\text{i} (n) r^\text{s} (n) } E^\text{i} ( r^\text{i}, \theta^\text{i}, \phi^\text{i} ) . 
\end{multline}
Here, $\omega_n$ represents the phase configuration of the $n$-th unit within the RIS. The electric field at the observation point is determined by the superposition of individual fields scattered by $N$ units. As
\begin{multline}\label{Model_SISO}
  E^\text{s} ( r^\text{s}, \theta^\text{s}, \phi^\text{s} ) = E^\text{i} ( r^\text{i}, \theta^\text{i}, \phi^\text{i} ) \sum_{n=1}^{N} \tau ( \theta^\text{s} (n), \phi^\text{s} (n) ; \theta^\text{i} (n), \phi^\text{i} (n) ) \\ 
  e^{j \omega_n} \frac{ e^{ - j 2 \pi r^\text{i} (n) / \lambda} e^{ - j 2 \pi r^\text{s} (n) / \lambda } }{ r^\text{i} (n)  r^\text{s} (n) }   . 
\end{multline}

When the RIS is exposed to multiple incident waves, the observed electric field is obtained through the superposition of individual fields
\begin{equation}\label{Model_MISO}
\begin{aligned}
  E^\text{s} ( r^\text{s}, \theta^\text{s}, \phi^\text{s} ) = 
  & \sum_{m=1}^{M} E^\text{i}_m ( r^\text{i}_m, \theta^\text{i}_m, \phi^\text{i}_m )  \\
  & \sum_{n=1}^{N} \tau ( \theta^\text{s} (n), \phi^\text{s} (n) ; \theta^\text{i}_m (n), \phi^\text{i}_m (n) ) \\ 
  & e^{j \omega_n} \frac{ e^{ - j 2 \pi r^\text{i}_m (n) / \lambda} e^{ - j 2 \pi r^\text{s} (n) / \lambda } }{ r^\text{i}_m (n)  r^\text{s} (n) }   . 
\end{aligned}
\end{equation}

Note that the input/output behavior expression \eqref{SISOBehavior} can be simplified by considering specific scenarios.

\begin{figure}[!htbp]
  \centering
  \includegraphics[width=0.8\linewidth]{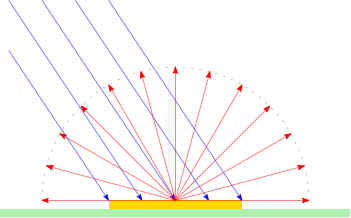}
  \caption{Incident EM waves uniformly scatter in all directions over the hemisphere of reflection.}
  \label{Isotropic}
\end{figure}

We initially explore the scenario where the RIS comprises isotropic units. In isotropic scattering, incident EM waves uniformly scatter in all directions over the hemisphere of reflection, as depicted in Fig.~\ref{Isotropic}, irrespective of the angle of incidence or observation. Consequently, the unit's scattering pattern is simplified to $\tau_n ( \theta^\text{s}, \phi^\text{s}; \theta^\text{i}, \phi^\text{i} ) = \tau$. The electric field at the observation point is given as
\begin{multline}\label{SISOBehavior2}
  E^\text{s} ( r^\text{s}, \theta^\text{s}, \phi^\text{s} ) \\
  = E^\text{i} ( r^\text{i}, \theta^\text{i}, \phi^\text{i} ) \sum_{n=1}^{N} \tau e^{j \omega_n} 
  \frac{ e^{ - j 2 \pi r^\text{i} (n) / \lambda} e^{ - j 2 \pi r^\text{s} (n) / \lambda } }{ r^\text{i} (n)  r^\text{s} (n) }   . 
\end{multline}

\begin{figure}[!htbp]
  \centering
  \includegraphics[width=0.8\linewidth]{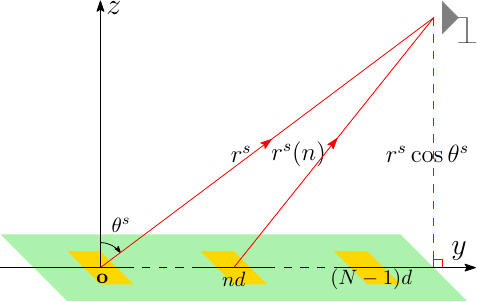}
  \caption{Geometry used for field calculations of a line source along the y-axis.}
  \label{FarField}
\end{figure}

\begin{figure*}[!htbp]
  \centering
  \includegraphics[width=0.65\textwidth]{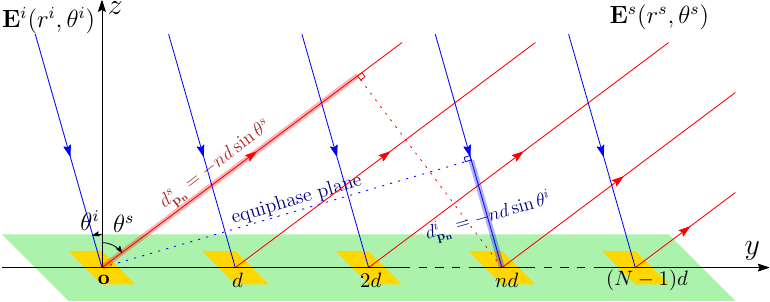}
  \caption{Phase discrepancy due to the interelement path length difference for linear RISs. The RIS is illuminated by uniform plane waves lying in the $yoz$ plane. The incident EM wave (in blue) is originating from $\theta^\text{i}$. The scattered electric field is observed at $(r^\text{s}, \theta^\text{s})$ (in red). The incident and scattered path length differences between the origin and the unit at $(n-1)d$ are $d_{\mathbf{p}_n}^\text{i} = - (n-1)d \sin \theta^\text {i} $ and $d_{\mathbf{p}_n}^\text{s} = - (n-1)d \sin \theta^\text s$, respectively.}
  \label{Fig:LinearRIS_1d}
\end{figure*}

Next, we will explore the scenario in the far-field regime, where the distances $r^\text{s} (n)$ and $r^\text{i} (n)$ exhibit linear relationships with $r^\text{s}$ and $r^\text{i}$, respectively. For the sake of clarity, we restrict our analysis to a uniform linear array. By considering the geometry depicted in Fig.~\ref{FarField}, we have
\begin{equation}
  \begin{aligned}
  r^\text{s} (n) = & \sqrt {(r^\text{s} \cos \theta^\text s )^2 + \left( r^\text{s} \sin \theta^\text s - n d \right)^2 } \\
  = & \sqrt {(r^\text{s} )^2 - 2 r^\text{s}  n d \sin \theta^\text s + \left( n d \right)^2 } \\
  = & r^\text{s} \sqrt { 1 - 2 \frac{ n d }{ r^\text{s} } \sin \theta^\text s + \left( \frac{ n d }{ r^\text{s} } \right)^2 },
  \end{aligned}
\end{equation}
where $d$ is the inter-unit spacing.
In the far-field regime, we assume  $ n d \ll r^\text{s}$, or equivalently $n d / r^\text{s} \ll 1$. To obtain approximate expressions for $r^\text{s} (n)$, we employ the approximation $\sqrt{ 1 + x } \approx 1 + x/2$. By neglecting high-order terms, we have 
\begin{equation}
  \begin{aligned}
  & \sqrt { 1 - 2 \frac{ n d }{ r^\text{s} } \sin \theta^\text{s} + \left( \frac{ n d }{ r^\text{s} } \right)^2 } \\
\approx & 1 - \frac{ n d }{ r^\text{s} } \sin \theta^\text{s} + \frac{1}{2} \left( \frac{ n d }{ r^\text{s} } \right)^2 
\approx 1 - \frac{ n d }{ r^\text{s} } \sin \theta^\text{s}.
  \end{aligned}
\end{equation}
Consequently, we arrive at
\begin{equation}
  r^\text{s} (n) \approx r^\text{s} - n d \sin \theta^\text{s}
\end{equation}
and
\begin{equation}
  r^\text{i} (n) \approx r^\text{i} - n d \sin \theta^\text{i} .
\end{equation}
These approximations are visually demonstrated in Fig.~\ref{Fig:LinearRIS_1d}.

Finally, the electric field at the observation point is reduced to
\begin{multline}
  E^\text{s} ( r^\text{s}, \theta^\text{s} ) \approx \tau \frac{ e^{ - j 2 \pi r^\text{i} / \lambda}  }{ r^\text{i} }  \frac{ e^{ - j 2 \pi r^\text{s} / \lambda } }{ r^\text{s} } E^\text{i} ( r^\text{i}, \theta^\text{i} ) \\
  \sum_{n=0}^{N-1} e^{j \omega_n} e^{ j 2 \pi n d \sin \theta^\text{i} / \lambda }  e^{ j 2 \pi n d \sin \theta^\text{s} / \lambda } .
\end{multline}

When the RIS is illuminated by multiple point sources lying in the far field, whose strengths are denoted by $ \{ E^\text{i} (r^\text{i}_m, \theta^\text{i}_m), m=1, \ldots, M\}$, the scattered electric field observed at $(r^\text{s}, \theta^\text{s})$ is given by the superposition of individual fields
\begin{multline}\label{E:ScatteringLinearRIS}
  E^\text{s} ( r^\text{s}, \theta^\text{s} ) \approx \tau \frac{ e^{ - j 2 \pi r^\text{s} / \lambda } }{ r^\text{s}  } \sum_{m=1}^{M} \frac{ e^{ - j 2 \pi r^\text{i}_m / \lambda } }{ r^\text{i}_m } E^\text{i}_m ( r^\text{i}_m, \theta^\text{i}_m ) \\
  \sum_{n=0}^{N-1} e^{j \omega_n} e^{ j 2 \pi n d \sin \theta^\text{i}_m / \lambda }  e^{ j 2 \pi n d \sin \theta^\text{s} / \lambda } .
\end{multline}

\begin{figure}[!htbp]
  \centering
  \includegraphics[width=0.9\linewidth]{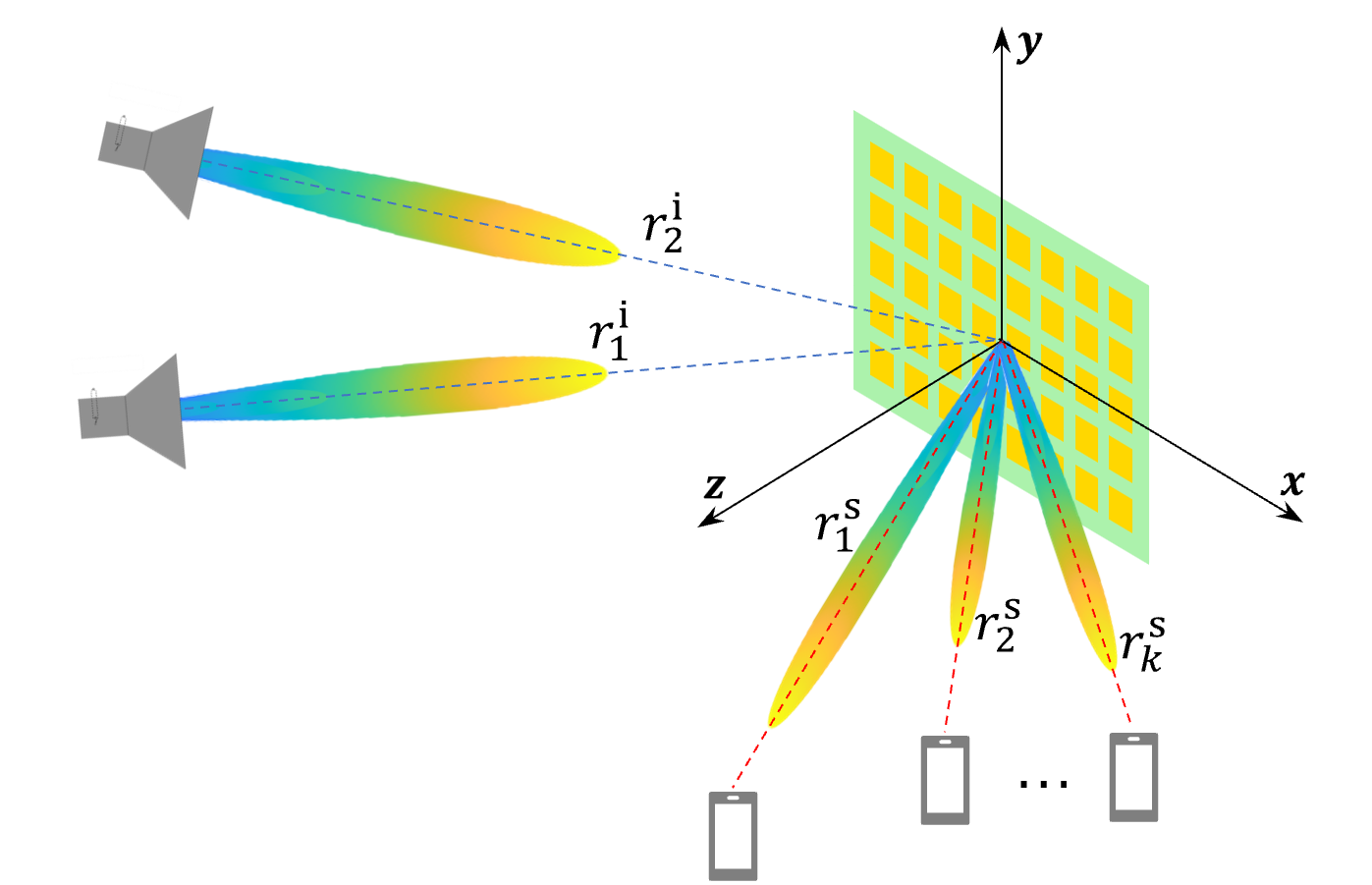}
  \caption{RIS-aided multiple-input-multiple-output communication.}
  \label{Fmulti}
\end{figure}

On the observation end, considering multiple points $\{ (r_1^\text{s}, \theta_{1}^\text{s}), \cdots, (r_T^\text{s}, \theta_{T}^\text{s}) \}$, as illustrated in Fig.~\ref{Fmulti}, we have the canonical linear representation \eqref{E:MIMO} to describe the input/output behaviors. The advantage lies in employing a simple system of linear equations to describe input/output behaviors, making it suitable for analyzing and optimizing the performance of RIS-aided systems.

\begin{figure*}[!htbp]
  \begin{equation}\label{E:MIMO}
    \begin{aligned}
      \begin{bmatrix}
        E^{s} (r_1^\text{s}, \theta_{1}^\text{s}) \\
        \vdots                           \\
        E^{s} (r_T^\text{s}, \theta_{T}^\text{s})
      \end{bmatrix}
      \approx & \tau 
      \begin{bmatrix}
        \frac{ e^{-j 2 \pi r_1^\text{s} / \lambda }}{ r_1^\text{s} } &        & 0 \\
                                                       & \ddots & \\
         0                                             &        & \frac{ e^{-j 2 \pi r_T^\text{s} / \lambda }}{ r_T^\text{s} }
      \end{bmatrix}
      \begin{bmatrix}
        1      & \cdots & e^{ j 2 \pi (N-1) d \sin \theta_1^\text{s} / \lambda } \\
        \vdots & \ddots & \vdots                                          \\
        1      & \cdots & e^{ j 2 \pi (N-1) d \sin \theta_T^\text{s} / \lambda } \\
      \end{bmatrix} 
      \begin{bmatrix}
            e^{j \omega_1} &        &  0 \\
                               & \ddots &    \\
            0                  &        & e^{j \omega_N}
      \end{bmatrix} \\
      & \begin{bmatrix}
        1                                                   & \cdots & 1                                                   \\
        \vdots                                              & \ddots & \vdots                                              \\
        e^{ j 2 \pi (N-1) d \sin \theta_{1}^\text{i} / \lambda } & \cdots & e^{ j 2 \pi (N-1) d \sin \theta_{M}^\text{i} / \lambda }
      \end{bmatrix} 
      \begin{bmatrix}
        \frac{ e^{-j 2 \pi r_1^\text{i} / \lambda }}{ r_1^\text{i} } &        &  0 \\
                            & \ddots &    \\
        0                   &        & \frac{ e^{-j 2 \pi r_M^\text{s} / \lambda }}{ r_M^\text{s} }
      \end{bmatrix}
      \begin{bmatrix}
        E^\text i (r_1^\text{i}, \theta_{1}^\text{i}) \\
        \vdots    \\
        E^\text i (r_M^\text{i}, \theta_{M}^\text{i})
      \end{bmatrix} .
    \end{aligned}
  \end{equation}
  \medskip
  \hrule
\end{figure*}

\subsection{Max-min Fair Beam Allocations Problem}

We explore a beam allocation strategy with the objective of maximizing the minimum power utility. This approach places significant emphasis on ensuring fairness among multiple users. Formally, this strategy can be expressed  as 
\begin{equation}\label{FB1}
  \max_{ \omega_1, \dots, \omega_{N} \in [0, 2\pi) } \min_{ k } \  \left \{ P^\text{s} ( r^\text{s}_k, \theta^\text{s}_k, \phi^\text{s}_k ), k = 1, \ldots, K \right \}
\end{equation}
Here $P^\text{s} (\cdot) = \lvert E^\text{s} (\cdot) \rvert^2$ represents the power at a specific location.

With the input/output models \eqref{SISOBehavior2}, \eqref{E:ScatteringLinearRIS}, and \eqref{E:MIMO}, we can derive representations of the received power under different scenarios. Similarly, we consider a uniform linear array as an example. By leveraging \eqref{E:ScatteringLinearRIS}, the received power can be expressed as
\begin{equation}
\begin{aligned}  
P^\text{s} ( r^\text{s}, \theta^\text{s} ) 
\approx & \left \lvert \frac{ \tau  }{ r^\text{s} } \sum_{m=1}^{M} \frac{ e^{ - j 2 \pi r^\text{i}_m / \lambda } }{ r^\text{i}_m } E^\text{i}_m ( r^\text{i}_m, \theta^\text{i}_m ) \right. \\
& \left. \sum_{n=0}^{N-1} e^{j \omega_n} e^{ j 2 \pi n d \sin \theta^\text{i}_m / \lambda }  e^{ j 2 \pi n d \sin \theta^\text{s} / \lambda } \right \rvert^2 \\
= & \left \lvert \frac{ \tau  }{ r^\text{s} } \sum_{n=0}^{N-1} e^{j \omega_n} e^{ j 2 \pi n d \sin \theta^\text{i}_m / \lambda }  e^{ j 2 \pi n d \sin \theta^\text{s} / \lambda } \right. \\
& \left. \sum_{m=1}^{M} \frac{ e^{ - j 2 \pi r^\text{i}_m / \lambda } }{ r^\text{i}_m } E^\text{i}_m ( r^\text{i}_m, \theta^\text{i}_m )\right \rvert^2 \\
= & \left \lvert \frac{ \tau  }{ r^\text{s} } \sum_{n=0}^{N-1} e^{j \omega_n} h_n ( \theta^\text{s} ) \right \rvert^2 .
\end{aligned}
\end{equation}
Here, $h_n ( \theta^\text{s} )$ is defined as the term 
\[
e^{ j 2 \pi n d \sin \theta^\text{i}_m / \lambda }  e^{ j 2 \pi n d \sin \theta^\text{s} / \lambda } \sum_{m=1}^{M} \frac{ e^{ - j 2 \pi r^\text{i}_m / \lambda } }{ r^\text{i}_m } E^\text{i}_m ( r^\text{i}_m, \theta^\text{i}_m ).
\]
If we define $\alpha(r^\text{s}) = \lvert \tau  / r^\text{s} \rvert^2$, $\mathbf{h} ( \theta^\text{s} ) = [h_1 ( \theta^\text{s} ), \cdots, h_N ( \theta^\text{s} )]^T$, and $\mathbf{w} = [e^{j \omega_1}, \cdots, e^{j \omega_N}]^H$, the received power can be represented as
\begin{equation}\label{E:Power}
  P^\text{s} ( r^\text{s}, \theta^\text{s} ) \approx \alpha(r^\text{s}) \mathbf{w}^H \mathbf{H} ( \theta^\text{s} ) \mathbf{w} ,
\end{equation}
where $\mathbf{H} ( \theta^\text{s} ) = \mathbf{h} ( \theta^\text{s} ) \mathbf{h} ( \theta^\text{s} ) ^H$ serves as a steering matrix.

\begin{remark}
   This expression of received power offers the advantage of decoupling the two essential factors in beamforming: distance and angle. The steering matrix $\mathbf{H} ( \theta^\text{s} )$ relies solely on the reflective angle $\theta^\text{s}$ when the incident waves are given and fixed. In contrast, $\alpha(r^\text{s})$ is independent of the angle $\theta^\text{s}$. Ideally, it is inversely proportional to the square of the distance. From another perspective, $\alpha(\cdot)$ serves as a weight, and other factors, such as user priority, can be incorporated.  It is a versatile parameter that can be utilized for power level tuning. By independently controlling $\alpha(\cdot)$ and $\mathbf{H} ( \theta^\text{s} )$, various beam redistributions functionalities can be implemented.
\end{remark}

Finally, we arrive at the Max-min expression for fair beam allocation through RISs
\begin{equation}\label{FB2}
\begin{aligned}
\text{(P1)} \ \max_{ \omega_1, \dots, \omega_{N} \in [0, 2\pi) } \min_{k} \  \left \{ \alpha( r^\text{s}_k )  \mathbf{w}^H \mathbf{H} ( \theta^\text{s}_k ) \mathbf{w} , k=1, \ldots, K \right \}.
\end{aligned}
\end{equation}
Here $\alpha( r^\text{s}_k ) > 0$ and $\mathbf{H} ( \theta^\text{s}_k )$ is a non-negative definite matrix with rank one for each $k$. In fact, problem (P1) is a special case of a general Max-min optimization problem involving quadratic forms
\begin{equation}\label{MM2}
\begin{aligned}
\text{(P0)} \ \max_{ \omega_1, \dots, \omega_{N} \in [0, 2\pi) } \min_{k} \ \left \{ \mathbf{w}^H \mathbf{H}_k \mathbf{w} , k=1, \ldots, K \right \} .
\end{aligned}
\end{equation}
Here $\mathbf{H}_k$ can be an arbitrary non-negative definite matrix. It is important to note that besides the optical model above, this quadratic problem form frequently emerges in statistical channel models and other research domains~\cite{xiong2022optimal}. We will develop methods effectively addressing this problem in the following Section. We emphasize once more that $\mathbf{w}$  resides in the union of tori, a highly non-convex set, rather than the Hilbert space $\mathbb{C}^N$.

\section{Moreau-Yosida Approximation-based Method}\label{Section3}
Optimization problems (P0) and (P1) belong to the class of semi-infinite Max-min problems. Our proposed method is related to MM approach, but with special techniques. We make use of Moreau-Yosida approximation of the maximum function $M(\mathbf{x})=\underset{k}{\max}\,x_k$, which can be explicitly written as \cite{zhang2008compensated,zhang2022tight}
\begin{equation}
\begin{aligned}
M_\mu(\mathbf{x})=&\underset{\mathbf{z}}{\min}\,M(\mathbf{z})+\mu\|\mathbf{z}-\mathbf{x}\|^2\\
=&\mu\|\mathbf{x}\|^2 - \frac{1}{4\mu}\|2\mu \mathbf{x}-P_\Delta(2\mu \mathbf{x})\|^2,
\end{aligned}
\end{equation}
where $\Delta=\{\mathbf{x}\in\mathbb{R}_+^n:\sum_k x_k \leq 1\}$ is the unit simplex, $P_\Delta$ is the projection onto $\Delta$, and $\mu$ is the regualarization parameter. It is well-known that Moreau-Yosida approximation is continuous differentiable with Lipschitz continuous gradient ($C^{1,1}$  smooth), and the Lipschitz constant of its gradient is at most $2\mu$ \cite{parikh2014proximal}. In our case, the gradient of $M_\mu(\mathbf{x})$ is given by
\begin{equation}\label{DxM}
D_x M_\mu(\mathbf{x})=P_\Delta(2\mu \mathbf{x}).
\end{equation}
It can be proved that \cite{zhang2008compensated,zhang2022tight}
\begin{equation}\label{eq:err}
-\frac{1}{4\mu}\leq M_\mu(\mathbf{x})-M(\mathbf{x})\leq 0.
\end{equation}

We denote ${f}_k(\mathbf{w})=\mathbf{w}^H \mathbf{H}_k \mathbf{w}$, then \eqref{MM2} becomes 
\begin{equation}\label{min-max}
\begin{aligned}
\max_{\omega_1, \dots, \omega_{N}} \min_{k}\,-{f}_k(\mathbf{w})=&-\min_{\omega_1, \dots, \omega_{N}} \max_{k}\,{f}_k(\mathbf{w})\\ =& -\min_{\omega_1, \dots, \omega_{N}}M(\mathbf{f}(\mathbf{w}))
\end{aligned}
\end{equation} 
Then we propose the smooth unconstrained optimization problem as
\begin{equation}\label{uncons}
\min_{\omega_1, \dots, \omega_{N}} M_\mu(\mathbf{f}(\mathbf{w})),
\end{equation}
where $\mathbf{f}(\mathbf{w})=(f_1(\mathbf{w}),\dots,f_K(\mathbf{w}))$. Problem \eqref{uncons} can be solved by gradient-based methods such as Nesterov's accelerated gradient descent method \cite{nesterov1998introductory,devolder2014first}. It is well-established that the gradient descent of the Moreau-Yosida approximation of a given function is equivalent to the proximal gradient algorithm for minimization of that function\cite{parikh2014proximal}. However, in \eqref{uncons} the objective function is the Moreau-Yosida approximation of the maximum function {\it composed with} a vector-valued function $\mathbf{f}(\mathbf{w})$, so the proximal gradient is not relevant.  We note that the gradient of $M_\mu(\mathbf{f}(\mathbf{w}))$ is given by \eqref{DxM} and the chain rule:
\begin{equation}
D_{\boldsymbol{\omega}} M_\mu(\mathbf{f}(\mathbf{w})) = D_{\boldsymbol{\omega}} \mathbf{f}(\mathbf{w})\cdot P_\Delta(2\mu \mathbf{f}(\mathbf{w})).
\end{equation}
Here $D_{\boldsymbol{\omega}} \mathbf{f}(\mathbf{w}) = (D_{\boldsymbol{\omega}} f_1(\mathbf{w}),\dots,D_{\boldsymbol{\omega}} f_K(\mathbf{w}))$, and 
\begin{equation}
D_{\boldsymbol{\omega}} f_k(\mathbf{w}) = -2\alpha_k\mathcal{\Im}(\bar{\mathbf{w}}\circ \mathbf{H}_k\mathbf{w}),
\end{equation}
where $\circ$ refers to the Hadamard product.
We denote 
\begin{equation}\label{eq:p}
P_\Delta(2\mu \mathbf{f}(\mathbf{w}))=\mathbf{p}(\mathbf{w})=(p_1(\mathbf{w}),\dots,p_K(\mathbf{w})),
\end{equation}
then 
\begin{equation}\label{eq:DMlambda}
D_{\boldsymbol{\omega}} M_\mu(\mathbf{f}(\mathbf{w})) = \sum_{p_k>0} p_k\,D_{\boldsymbol{\omega}}f_k(\mathbf{w}).
\end{equation} 

An important necessary condition for the optimization problems is the first order optimal condition \cite{nocedal1999numerical}. The first order optimal condition for \eqref{min-max} is that for minimizer $\boldsymbol{\omega}^* = [\omega_1^*,\cdots,\omega_N^*]^T $ of \eqref{min-max}, there exists a non-negative non-zero vector $\mathbf{p}^*=[p_1^*,\dots,p_K^*]^T$ such that
\begin{subequations}\label{eq:KKT}
\begin{align}
&\sum_{k=1}^n p_k^*\,D_{\boldsymbol{\omega}}f_k(\mathbf{w}^*)=0;\\
&\max_{k}\,{f}_k({\mathbf{w}^*}) = f_l({\mathbf{w}^*}) , \text{ for every }p_l^*>0, \label{eq:compslack}
\end{align}
\end{subequations}
where $\mathbf{w}^*=(e^{j\boldsymbol{\omega}^*})^H$.

The following theorem shows that the minimizer of \eqref{uncons} approximately satisfies the first order optimal condition \eqref{eq:KKT} for \eqref{min-max}  provided that $\mu$ is sufficiently large. 

\begin{thm}\label{thm:KKT2}
Suppose $\{f_k(\mathbf{w})\}_{k=1}^K$ are differentiable functions bounded from below. Then for every $\epsilon>0$, if $\mu>\frac{1}{2\epsilon}$, then there exists some non-negative non-zero vector $\hat{\mathbf{p}}=[\hat{p}_1,\dots,\hat{p}_K]^T$ such that the minimizer $\hat{\boldsymbol{\omega}}$ of \eqref{uncons} satisfies
\begin{subequations}\label{eq:approx_KKT}
\begin{align}
&\sum_{k=1}^n \hat{p}_k\,D_{\boldsymbol{\omega}}f_k(\hat{\mathbf{w}})=0;\label{eq:optimal}\\
&|f_l(\hat{\mathbf{w}}) - \max_{k}\,{f}_k(\hat{\mathbf{w}})|  <\epsilon, \text{ for every }\hat{p}_l>0, \label{eq:approx_compslack}
\end{align}
\end{subequations}
where $\hat{\mathbf{w}}=(e^{j\hat{\boldsymbol{\omega}}})^H$.

\end{thm}
\begin{proof}
Condition \eqref{eq:optimal} follows from the optimality condition of \eqref{uncons}, which is defined in \eqref{eq:DMlambda}. By \eqref{eq:p}, $(\hat{p}_1,\dots,\hat{p}_K)\in\Delta$, hence it is nonzero and non-negative.

Without loss of generality we assume that $f_1(\hat{\mathbf{w}})\geq f_2(\hat{\mathbf{w}})\geq \dots\geq f_K(\hat{\mathbf{w}})$, then by \cite{zhang2022tight} (Theorem 2.2) we have $\hat{p}_1\geq \hat{p}_2\geq\dots\geq\hat{p}_K$. Moreover, if $f_1(\hat{\mathbf{w}})-f_l(\hat{\mathbf{w}})\geq \frac{1}{2\mu}$, then $\hat{p}_k=0$. Conversely, if $\hat{p}_l>0$, then $f_1(\hat{\mathbf{w}})-f_l(\hat{\mathbf{w}})< \frac{1}{2\mu}$, that is exactly \eqref{eq:approx_compslack}. This completes the proof.
\end{proof}
\begin{remark} We call the solution $\hat{\boldsymbol{\omega}}$ satisfying the condition \eqref{eq:approx_KKT} a $\epsilon$-approximate the first order optimal point for the original problem \eqref{min-max}.
\end{remark}

\begin{algorithm}
\caption{Moreau-Yosida approximation-based (MA) method for Max-min problem~\eqref{MM2}}
\label{alg-1.0}
\begin{algorithmic}[1]
\Procedure{Max-Min}{$\mathbf{f}, \epsilon$}\Comment{solve \eqref{min-max} given tolerance $\epsilon$}
   \State initialize $\mu_0=10^{-3}, \text{gap}=1$
   \While{$\text{gap}>\epsilon $}
      \State solve \eqref{uncons} for ${\boldsymbol{\omega}}$ with current $\mu$
      \State $\mathbf{w} \leftarrow (e^{j\boldsymbol{\omega}})^H$ 
      \State calculate $\mathbf{p}$ by \eqref{eq:p}
      \State $I\gets \{k: p_k>0\}$
      \State $\text{gap}\gets \underset{k\in I}{\max}\,f_k(\mathbf{w})-\underset{k\in I}{\min}\,f_k(\mathbf{w})$
      \If {$\text{gap}>0$}
      		\State $\mu \leftarrow \max\{\frac{1}{2\cdot\text{gap}}, 2\mu\}$
      \EndIf
   \EndWhile\label{euclidendwhile}
   \State \textbf{return} $\mathbf{w}$
   \EndProcedure
\end{algorithmic} 
\end{algorithm}


Based on the above theorem, we propose an algorithm that iteratively solves \eqref{uncons} for increasing $\mu$ until the first order optimal condition is satisfied within user-defined tolerance $\epsilon$ as in \eqref{eq:approx_compslack}. The algorithm to solve the Max-min problem can be summarised in Algorithm \ref{alg-1.0}. 

The reason that we do not choose sufficiently large $\mu$ in the beginning is that the step size of the gradient-based algorithms is inversely proportional to the Lipschitz constant of the gradient of the objective function. As mentioned before, the Lipschitz constant of the gradient of Moreau-Yosida approximation $M_\mu(x)$ is proportional to $\mu$. To avoid small step sizes we begin with a relatively small $\mu$ and increase it adaptively until the first order optimal condition is satisfied within given tolerance $\epsilon$.

Note that the total number of outer iterations in Algorithm~\ref{alg-1.0} is at most $\log_2(\epsilon/\mu_0)$, where $\mu_0=10^{-3}$ in our experiments. The following theorem gives the error estimate of the proposed algorithm. 
\begin{thm}\label{thm:error_est}
Suppose minimizer of \eqref{min-max} is $\boldsymbol{\omega}^*$, then the minimizer $\hat{\boldsymbol{\omega}}$ of \eqref{uncons} satisfies
\begin{equation}\label{eq:error_bnd}
-\frac{1}{4\mu}\leq M(\mathbf{f}(\mathbf{w}^*))-M(\mathbf{f}(\hat{\mathbf{w}}))\leq 0,
\end{equation}
where $\mathbf{w}^*=(e^{j\boldsymbol{\omega}^*})^H$ and $\hat{\mathbf{w}}=(e^{j\hat{\boldsymbol{\omega}}})^H$.
\end{thm}

\begin{proof}
By \eqref{eq:err} we have 
\begin{subequations}
\begin{align}
-\frac{1}{4\mu}\leq M_\mu(\mathbf{f}(\hat{\mathbf{w}}))-M(\mathbf{f}(\hat{\mathbf{w})})\leq 0,\\
0\leq M(\mathbf{f}({\mathbf{w}}^*))-M_\mu(\mathbf{f}({\mathbf{w}}^*))\leq \frac{1}{4\mu}.
\end{align}
\end{subequations}
And by optimality of $\hat{\boldsymbol{\omega}}$ in \eqref{uncons},
\begin{equation}
0\leq M_\mu(\mathbf{f}({\mathbf{w}}^*)) - M_\mu(\mathbf{f}(\hat{\mathbf{w}})).
\end{equation}
Adding these inequalities together we obtain
\begin{equation}
-\frac{1}{4\mu}\leq M(\mathbf{f}(\mathbf{w}^*))-M(\mathbf{f}(\hat{\mathbf{w}})),
\end{equation}
and the right half of \eqref{eq:error_bnd} follows from the optimality of ${\boldsymbol{\omega}}^*$ in \eqref{min-max}. This completes the proof.
\end{proof}


%
%
%
%

 \section{Flexible beam allocation functionalities}\label{Section4}

To evaluate the effectiveness of Max-min-based beam allocation approach, we conduct a series of numerical experiments. we initially validate various beam allocation functions of RISs, namely, beam splitting, fair beam allocation, and wide-beam generation. Subsequently, we assess the power level controlling and introduce the weight assignment scheme in dynamic changing environments.

\subsection{Beam-Spliting}

We initially evaluate the beam-splitting capability of RISs. In this functionality, a single incident wave interacts with the RIS, resulting in the generation of multiple reflected waves. The requirement is that the direction of these reflected waves can be arbitrarily controlled. This functionality is essential for designing flexible beams for the detection of multiple targets.

The employed RIS comprises $32\times32$ units. Suppose an incident wave originates from $(5 \ \text{m}, 0^\circ, 0^\circ)$. We utilize the model given in expression \eqref{Model_SISO} to calculate the steering matrix $\mathbf{H} ( \theta^\text{s} )$. The anticipated outcome involves the RIS splitting the incident wave into three directions: $(\theta^\text{s}_1,\phi^\text {s}_1) = (30^\circ,180^\circ)$, $(\theta^\text{s}_2,\phi^\text {s}_2) = (45^\circ,0^\circ)$, and $(\theta^\text{s}_3,\phi^\text {s}_3) = (60^\circ,0^\circ)$. For better illustration, we set the distance values of $\{ r^\text{s}_k, k = 1, 2, 3 \}$ to be identical.

\begin{figure}[!htbp]
  \centering
  \subfigure[]{
  \label{beamsplit1}
  \includegraphics[width=0.75\linewidth]{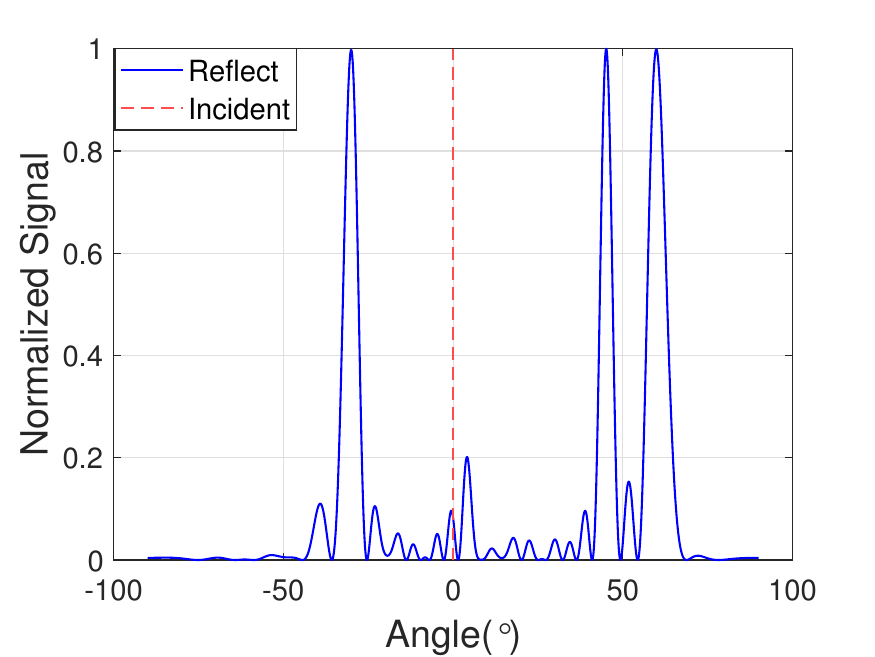}}
  \subfigure[]{
  \label{beamsplitpolar}
  \includegraphics[width=0.75\linewidth]{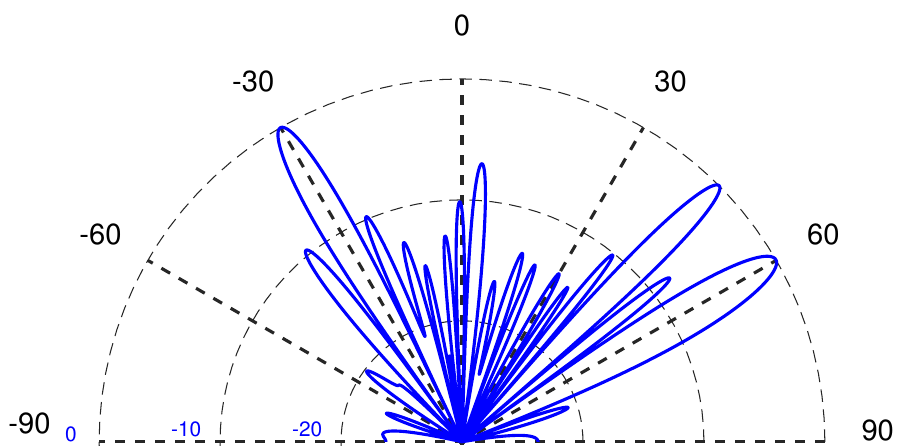}}
  \caption{Illustration of beam split in a normalized radiation beam pattern. The incident signal from $(\theta^\text{i},\phi^\text{i}) = (0^\circ,0^\circ)$, and reflect direction are set to be $(\theta^\text{s},\phi^\text{s}) = (30^\circ,180^\circ), (45^\circ,0^\circ), (60^\circ,0^\circ)$, respectively. (a) radiation beam pattern. (b) radiation beam pattern in polar plot}
  \label{beamsplit}
\end{figure}

From Fig.~\ref{beamsplit}, it is evident that a single incident signal, after reflection by properly configured RIS, is divided into three signals of equal power. It's noteworthy that achieving this function through traditional phase compensation configuration is impractical due to the anomalous reflection phenomenon, as previously clarified.


\subsection{Fair Beam Allocation}

In contrast to the beam-splitting functionality, which involves a single incident wave splitting into multiple reflected waves, fair beam allocation focuses on redistributing multiple incident waves across several directions. In this function, the RIS initially collects all incident power and redirects it toward the intended targets. This functionality is vital for enhancing capacity and expanding coverage to a broader range of locations.

For this validation, we assume four incident signals as parameterized in TABLE.~\ref{common}, and utilize the model given in \eqref{Model_MISO} to calculate the steering matrix $\mathbf{H} ( \theta^\text{s} )$. 
To ensure a comprehensive demonstration, we assess the performance of fair beam allocations with configurations of 1, 4, and 10 users (observation points). Similarly, the values of $\{ r^\text{s}_k, k=1, \ldots, K\}$ are set to be identical. 

\begin{figure}[!htbp]
  \centering
  \subfigure[]{
  \label{1polar}
  \includegraphics[width=0.7\linewidth]{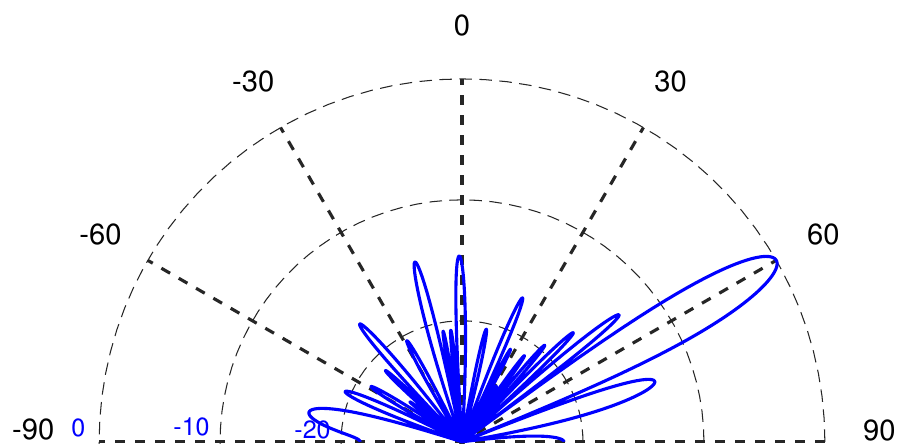}}
  \subfigure[]{
  \label{4polar}
  \includegraphics[width=0.7\linewidth]{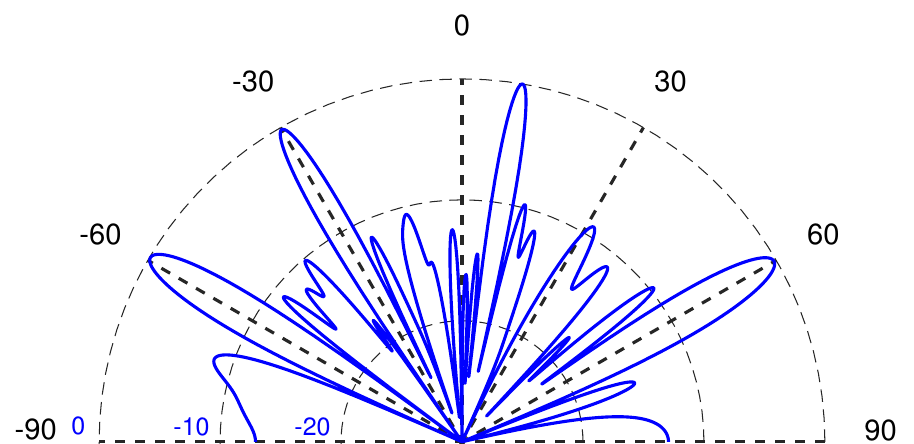}}
  \subfigure[]{
  \label{10polar}
  \includegraphics[width=.7\linewidth]{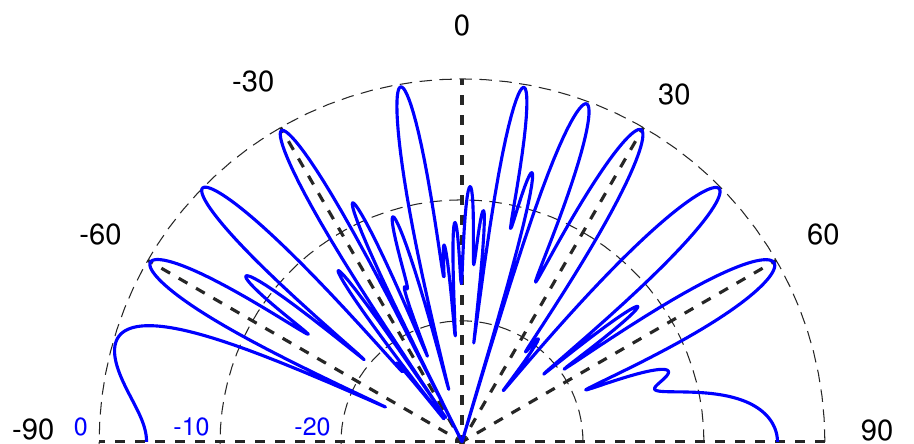}}
  \caption{Illustration of normalized radiation beam pattern of RIS, represented in polar plots. (a) The user direction is $(\theta^\text{s},\phi^\text{s})=(60^{\circ},0^{\circ})$. (b) The users directions are $(\theta^\text s,\phi^\text s)=(60^{\circ},180^{\circ}),(30^{\circ},180^{\circ}),(10^{\circ},0^{\circ}),(60^{\circ},0^{\circ})$. (c) The users directions are $(\theta^\text s,\phi^\text s)=(75^{\circ},180^{\circ}),(60^{\circ},180^{\circ}),(45^{\circ},180^{\circ}),(30^{\circ},180^{\circ}),(10^{\circ},180^{\circ}),(10^{\circ},0^{\circ})$.}
  \label{polar}
\end{figure}

The polarization plot results are depicted in Fig.~\ref{polar}. It illustrates a balanced distribution of directed power, with the beam directions aligned to the preset parameters. In Fig.~\ref{1polar}, we implement the beam focus function, which gathers all incident power and concentrates it in a single direction. This fundamental functionality plays an essential role in wireless energy harvesting.

Furthermore, it can be observed that, even without actively controlling the level of side lobes, the main lobe directed towards the user exhibits a power approximately 10 dB higher in all three setups. It can be envisioned that if there are means to control side lobes, the performance will be significantly enhanced. Achieving this improvement may involve introducing constraints within the fair beam allocation framework, which remains an area of ongoing research.


\subsection{Wide-beam Generation}

\begin{figure}[!htbp]
  \centering
  \subfigure[]{
  \label{widebeampolar}
  \includegraphics[width=0.75\linewidth]{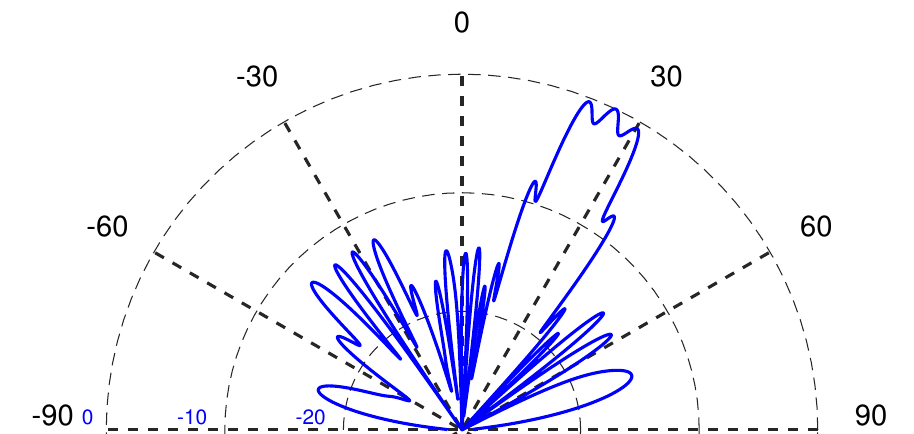}}
  \subfigure[]{
  \label{widebeam}
  \includegraphics[width=.5\linewidth]{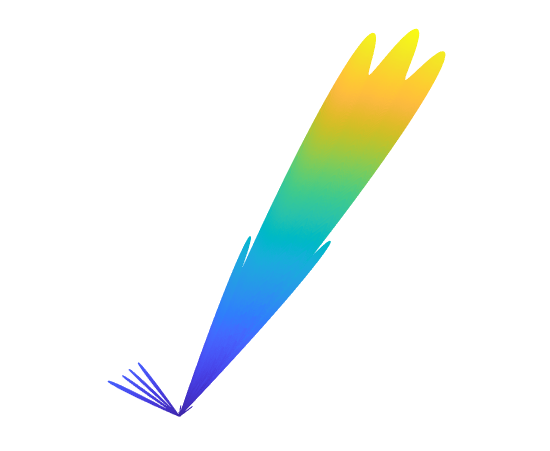}}
  \caption{Wide beam can be designed by synthesizing multiple sub-beams. (a) Radiation beam pattern in polar plot. (b) 3D scattered power pattern.}
  \label{wide-beam}
\end{figure}

In addition to designing multiple beams to serve users at various positions, the proposed approach can be employed to synthesize multiple sub-beams into a wide beam - a desirable functionality. Wide beams offer advantages in ensuring signal coverage and establishing reliable links across a broad range of areas. To achieve this function with the fair beam allocation framework, a series of closely spaced anchors are strategically selected. These sub-beams are automatically synthesized into a wide beam. An example is illustrated in Fig.~\ref{wide-beam}, where the wide beam is formed through the combination of 21 sub-beams ranging from elevation angle $20^\circ$ to $32^\circ$.

\subsection{Power Level Control}


In addition to uniform redistribution in the angular domain, the proposed fair beam allocation framework allows for the implementation of non-uniform redistribution. This power level control can be achieved by adjusting the weights $\{ \alpha(r^\text{s}_k) \}$ in problem (P1). We conduct several experiments to validate this non-uniform redistribution functionality. 

\begin{figure}[!htbp]
  \centering
  \subfigure[]{
  \label{5users}
  \includegraphics[width=.58\linewidth]{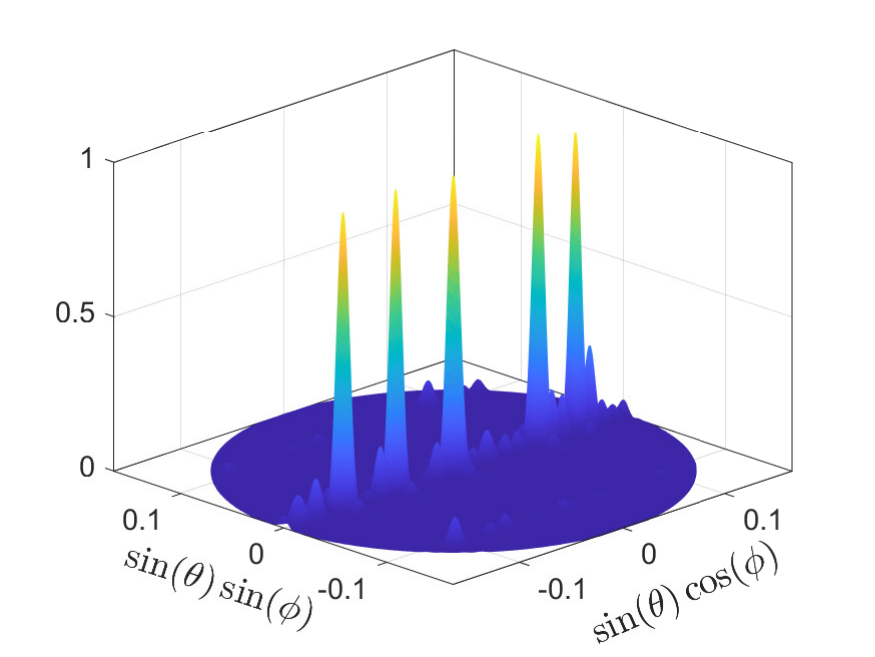}}
  \subfigure[]{
  \label{5users3D}
  \includegraphics[width=.38\linewidth]{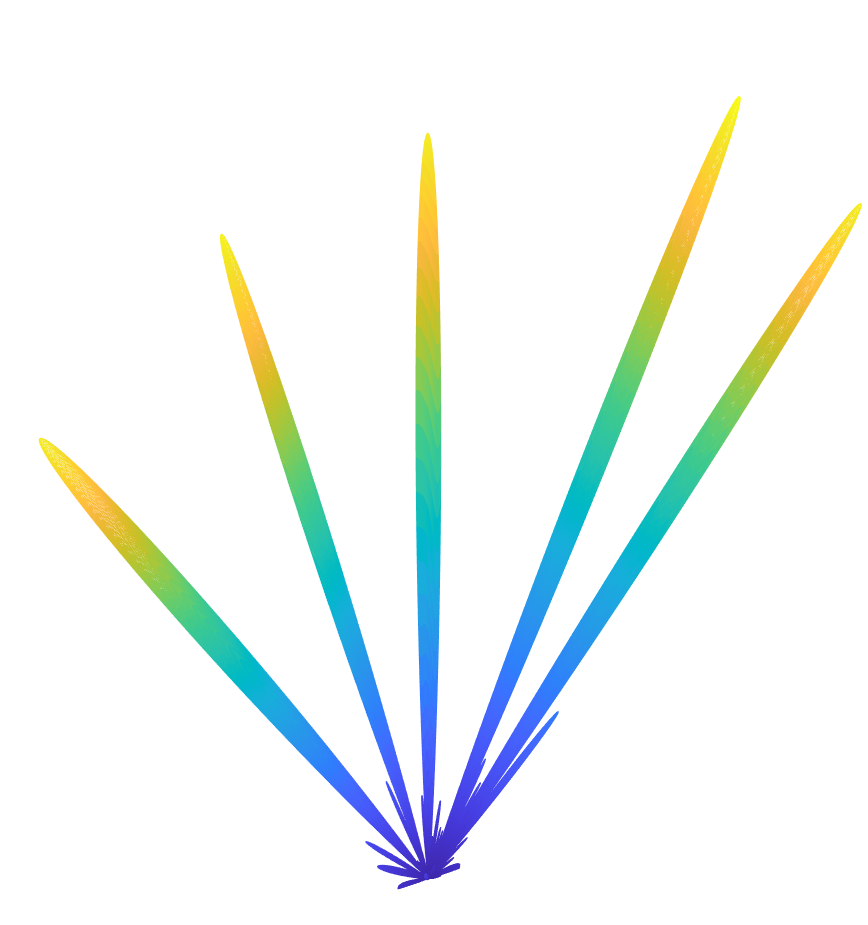}}
  \caption{Beamform design for five UEs with equal weight. UEs directions are $(0^\circ,0^\circ)$, $(20^\circ,180^\circ)$, $(30^\circ,0^\circ)$, $(40^\circ,180^\circ)$, $(45^\circ,0^\circ)$, respectively. The weights $\alpha(r^\text{s}_1), \alpha(r^\text{s}_2), \alpha(r^\text{s}_3), \alpha(r^\text{s}_4), \alpha(r^\text{s}_5)$ are set as $1/r^2, 1/r^2, 1/r^2, 1/r^2, 1/r^2$. (a) The normalized scattered power. (b) The 3D scattered power.}
  \label{10user}
\end{figure}

We initially employ reflective beam patterns to illustrate the impact of the weights. For comparison, we conduct three experiments with user equipments (UEs) located at an equal distance of $r=30\ \text{m}$. The reflective patterns corresponding to different weights are plotted. In the first experiment, we set $\alpha(r^\text{s}_1), \alpha(r^\text{s}_2), \alpha(r^\text{s}_3), \alpha(r^\text{s}_4), \alpha(r^\text{s}_5)$ as $1/r^2, 1/r^2, 1/r^2, 1/r^2, 1/r^2$. The resulting configurations ensure fair beam allocations at these anchor points, as shown in Figs.~\ref{10user}.



\begin{figure}[!htbp]
  \centering
  \subfigure[]{
  \label{Ntenusers}
  \includegraphics[width=.6\linewidth]{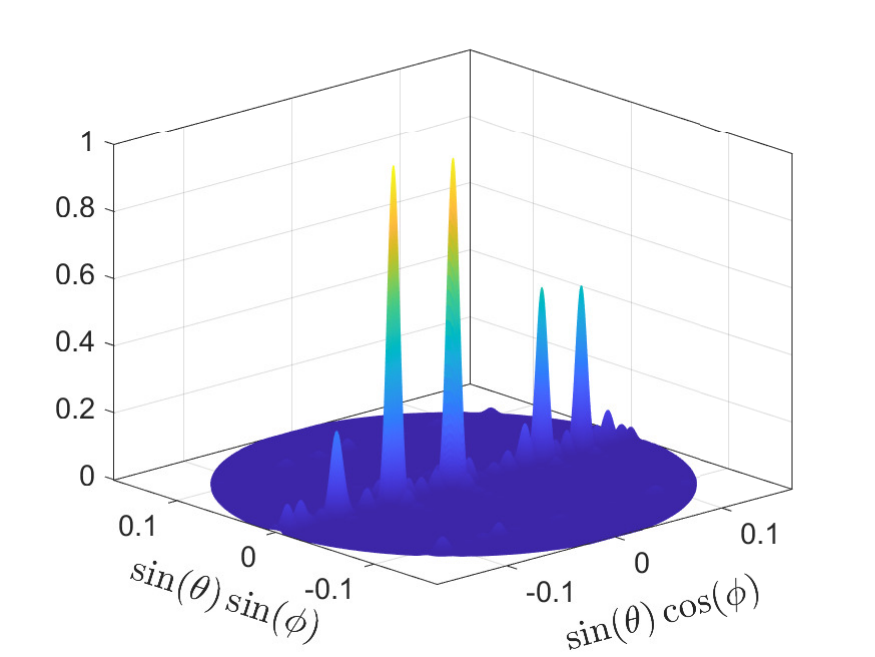}}
  \subfigure[]{
  \label{3Dtenusers}
  \includegraphics[width=.36\linewidth]{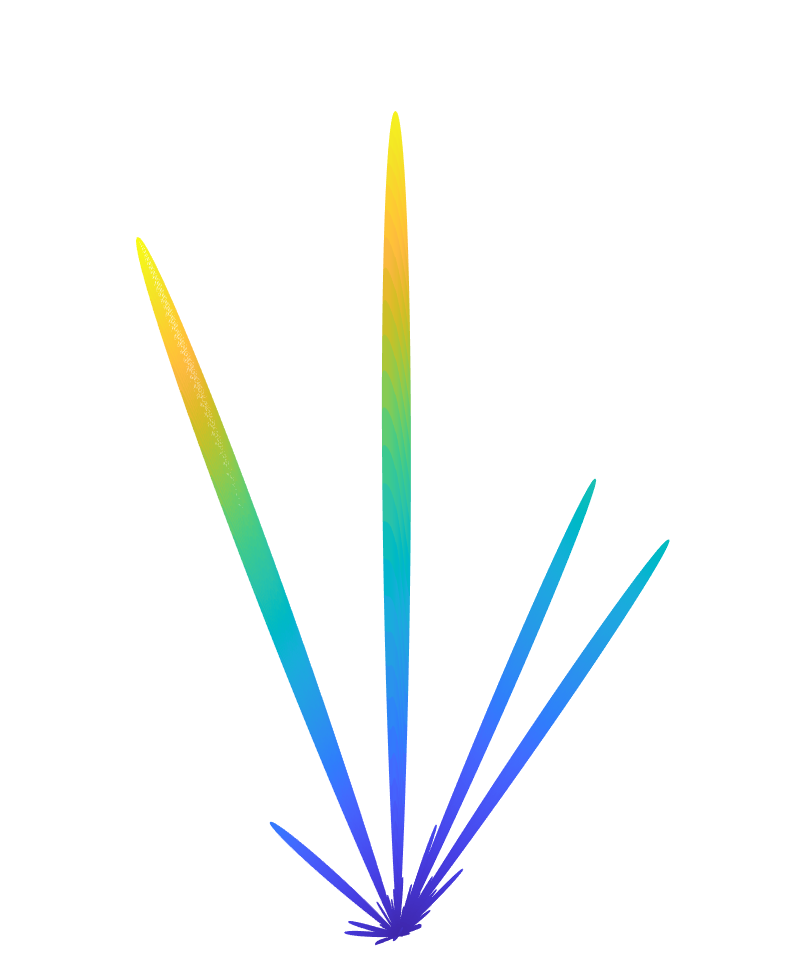}}
  \caption{Beamform design for five UEs with directions $(0^\circ,0^\circ)$, $(20^\circ,180^\circ)$, $(30^\circ,0^\circ)$, $(40^\circ,180^\circ)$, $(45^\circ,0^\circ)$, respectively. The weights are set as $1/4r^2, 1/4r^2, 1/2r^2, 1/2r^2, 1/r^2$. (a) The normalized scattered power. (b) The 3D scattered power.}
  \label{3user}
\end{figure}

\begin{figure}[!htbp]
\centering
\includegraphics[width=1\linewidth]{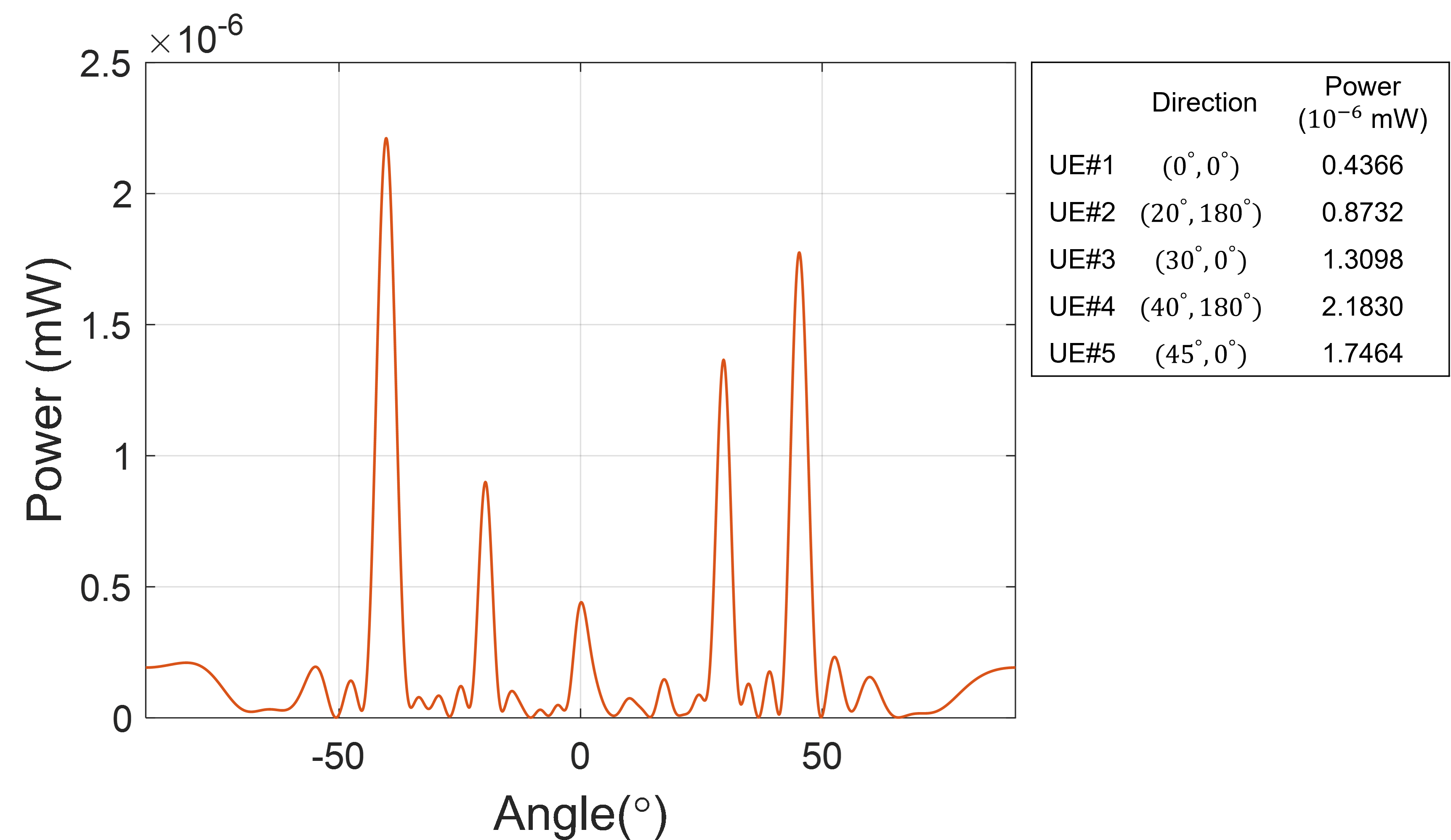}
\caption{RIS radiation beam pattern. Five UEs with directions $(0^\circ,0^\circ)$, $(20^\circ,180^\circ)$, $(30^\circ,0^\circ)$, $(40^\circ,180^\circ)$, $(45^\circ,0^\circ)$, respectively. The weights are set as $1/r^2, 1/2r^2, 1/3r^2, 1/5r^2, 1/4r^2$.}
\label{beam12345}
\end{figure}


Then, we assign the users respective weights of $1/4r^2, 1/4r^2, 1/2r^2, 1/2r^2, 1/r^2$. These weights are chosen to achieve a desired received power ratio of 4:4:2:2:1. The observed results, as shown in Fig.~\ref{3user}, manifest a signal power distribution consistent with the anticipated arrangement. The same conclusion can be summarized through the power curves in Fig.~\ref{beam12345}. The received powers at the five UEs align with the expected proportions of 1:2:3:5:4. These findings affirm the efficacy of the proposed methodology in governing the distribution of received signal power among users in RIS-aided communications.



To comprehensively illustrate the impact of weights, we simulate a more realistic scenario in which a user (UE\#5) dynamically moves away. We set UE\#5 to move from a distance of 10 m to 50 m. Meanwhile, UE\#1 to UE\#4 are fixed at different directions with distances to the RIS as 15 m, 12 m, 13 m, and 15 m, respectively. The received signal powers of all UEs are recorded throughout the movement of UE\#5. 

\begin{figure*}[!htbp]
  \centering
  \subfigure[]{
  \label{Ratio1111}
  \includegraphics[width=0.3\linewidth]{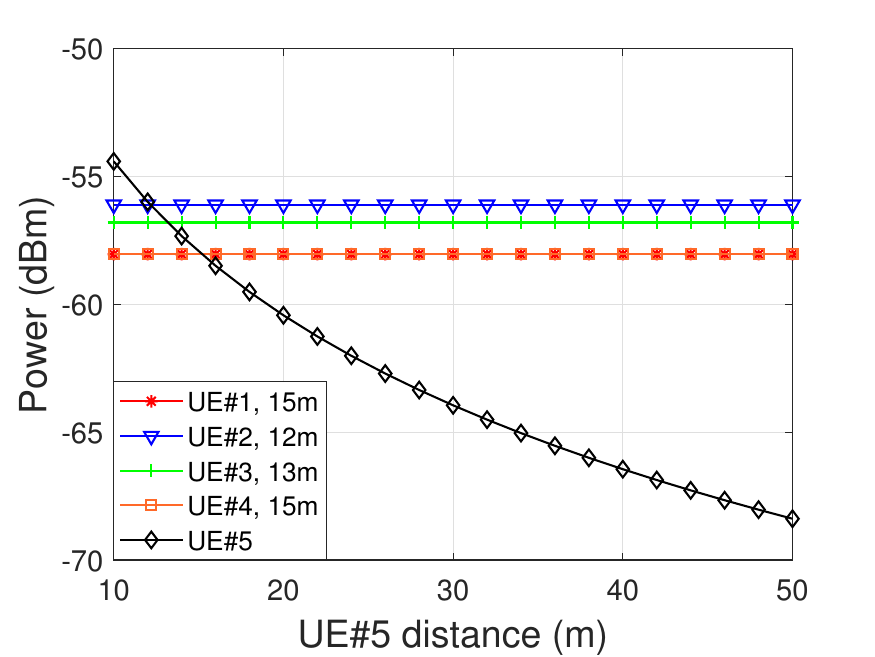}}
  \subfigure[]{
  \label{Ratio1111-05}
  \includegraphics[width=.3\linewidth]{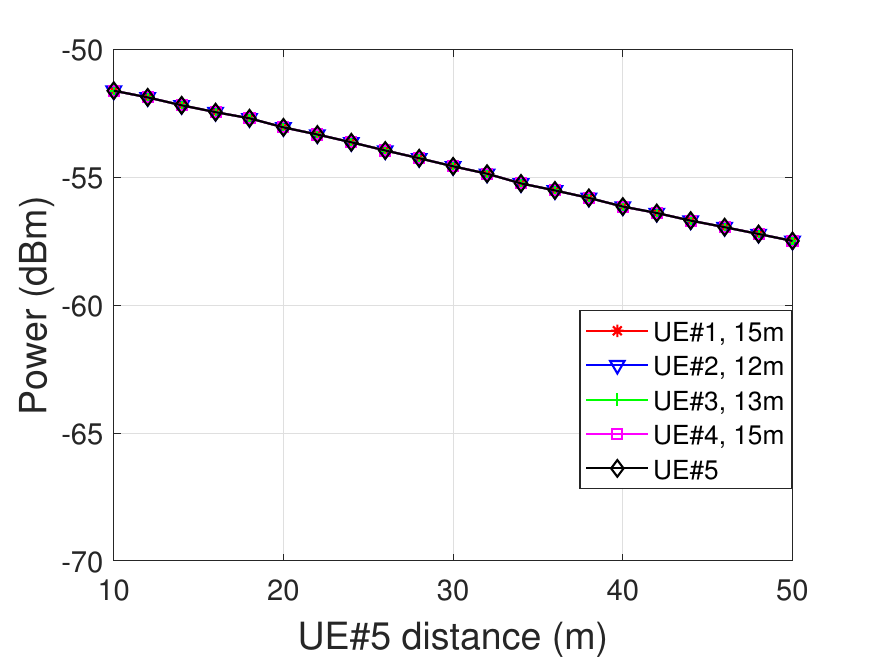}}
  \subfigure[]{
  \label{Ratio1111-01}
  \includegraphics[width=.3\linewidth]{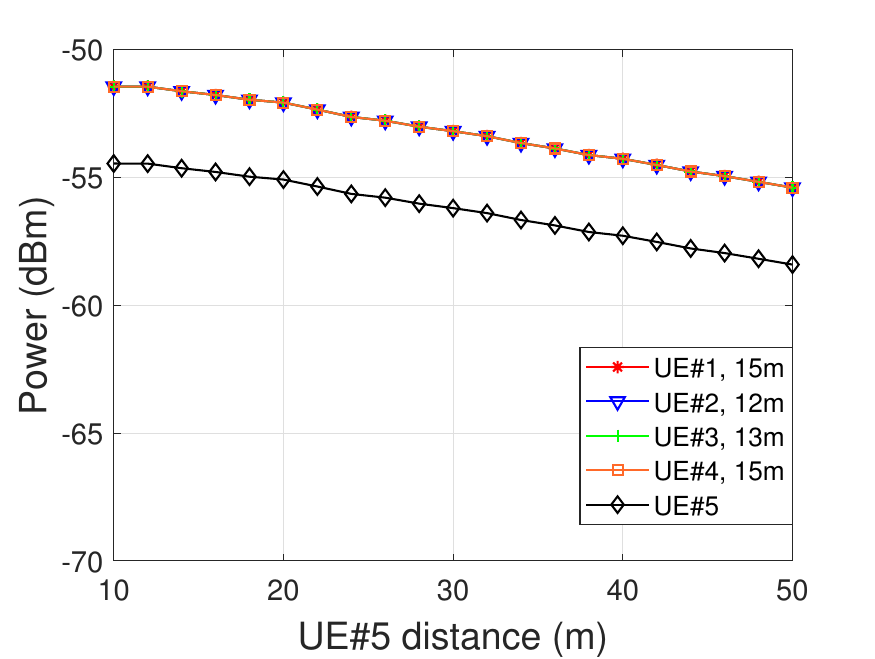}}
  \caption{The received signal powers while incrementally increasing the distance of UE\#5 to the RIS, the other UE positions are fixed. The weights $\alpha(r^\text{s}_k)$ from UE\#1 to UE\#5 are set as (a) 1, 1, 1, 1, 1. (b) $\frac{1}{(r_1^\text{s})^2}$, $\frac{1}{(r_2^\text{s})^2}$, $\frac{1}{(r_3^\text{s})^2}$, $\frac{1}{(r_4^\text{s})^2}$, $\frac{1}{(r_5^\text{s})^2}$. (c) $\frac{1}{(r_1^\text{s})^2}$, $\frac{1}{(r_2^\text{s})^2}$, $\frac{1}{(r_3^\text{s})^2}$, $\frac{1}{(r_4^\text{s})^2}$, $\frac{2}{(r_5^\text{s})^2}$, respectively.}
  \label{5user}
\end{figure*}

We initially explore a scenario with the absence of information on distance and movement, the default weight configuration $\alpha(r^\text{s}_1), \alpha(r^\text{s}_2), \alpha(r^\text{s}_3), \alpha(r^\text{s}_4), \alpha(r^\text{s}_5)$ as $1, 1, 1, 1, 1$. The corresponding received power is illustrated in Fig.~\ref{Ratio1111}. There is a rapid decline in received power for UE\#5. Specifically, when UE\#5 reaches the end at about 50 m, the rapid decrease in received power reaches 14 dB. In contrast, the other four UEs do not experience any power loss.

This situation is not desired and can be improved by adjusting the weight assignment strategy. As weights are strategically set as $\frac{1}{(r_1^\text{s})^2}$, $\frac{1}{(r_2^\text{s})^2}$, $\frac{1}{(r_3^\text{s})^2}$, $\frac{1}{(r_4^\text{s})^2}$, $\frac{1}{(r_5^\text{s})^2}$, the results in Fig.~\ref{Ratio1111-05} indicate that the received power for all UEs remains constant as UE\#5 moves. However, there is evident power loss for the other four UEs, attributed to compensating for UE\#5. Moreover, if we augment the weight of UE\#5 to $\frac{2}{(r_5^\text{s})^2}$, the declining trend in received power for all UEs is mitigated with the movement of UE\#5, as demonstrated in Fig~\ref{Ratio1111-01}. 

The appropriate distribution of weights is pivotal in achieving power-level control functionality. Moreover, in evolving communication environments, employing a dynamic weight strategy and considering the trade-off between signal power and change is imperative.

%
%

\section{Performance Evaluation and Comparison}\label{Section5}

After exploring various beam allocation functionalities, we proceed to assess the efficacy and efficiency of the proposed MA algorithm. We conduct evaluations of dual numerical simulations and real-world field trials. 
In the simulations, we compare the received power and execution-time performance of the proposed algorithm with two popular methods, namely Fminimax~\cite{fminmax} and QuantRand~\cite{subhash2023max}. 
Fminimax is the MATLAB implementation of the min-max solver based on the goal attainment method~\cite{brayton1979new,gembicki1974vector}. QuantRand is a quantized version of random coordinate descent, which is a greedy algorithm. 
Throughout the comparison, four incident signal sources are assumed to illuminate the RIS. The general parameter setting is illustrated in Table \ref{common}, if not emphasized.
In field trials, the 2-bit RIS prototypes operating at a frequency of 3.4 GHz are employed to evaluate the beamforming functionality.


\begin{table}[!htbp]
\caption{General Simulation Setup Parameters} \label{common}
\centering
\setlength{\tabcolsep}{1.2mm}
\setstretch{1.2} 
\begin{tabular}{ccc}
\toprule[1.5pt]
Description &  Symbol & Value  \\
\midrule
Operating frequency & $f$    & 3.4 GHz \\
Wavelength   &$\lambda$  & 0.088 m\\
Number of units     &$N$  & $32\times32$ \\
Inter-unit spacing  &$d $  &  0.044 m  \\
\multirowcell{2}{Number of incident signal \\(or transmitters, Tx)   }&\multirowcell{2}{$M$}    & \multirowcell{2}{4}  \\ \\
Tx\#1 position  &($r^\text{i}_1,\theta^{\text i}_1, \phi^{\text i}_1$) &(5 m, $0^{\circ},0^{\circ}$)\\ 
Tx\#2 position   &($r^\text{i}_2,\theta^{\text i}_2, \phi^{\text i}_2$) & (6 m, $40^{\circ},90^{\circ}$)\\
Tx\#3 position   &($r^\text{i}_3,\theta^{\text i}_3, \phi^{\text i}_3$) &  (8 m, $ 40^{\circ},180^{\circ}$)\\
Tx\#4 position   &($r^{\text i}_4,\theta^{\text i}_4, \phi^{\text i}_4$) &(4 m, $60^{\circ},0^{\circ}$)\\
Transmit signal power& $P_{\rm t}$   & 40 dBm \\
Number of users  &$K$  & 10\\ 
Distance from user to RIS  &$r^{\text s}$ &30 m\\ 
\bottomrule[1.5pt] 
\label{Common}
\end{tabular}
\end{table}

\begin{figure}[!htbp]
  \centering
  \subfigure[]{
  \label{beam111-3}
  \includegraphics[width=0.75\linewidth]{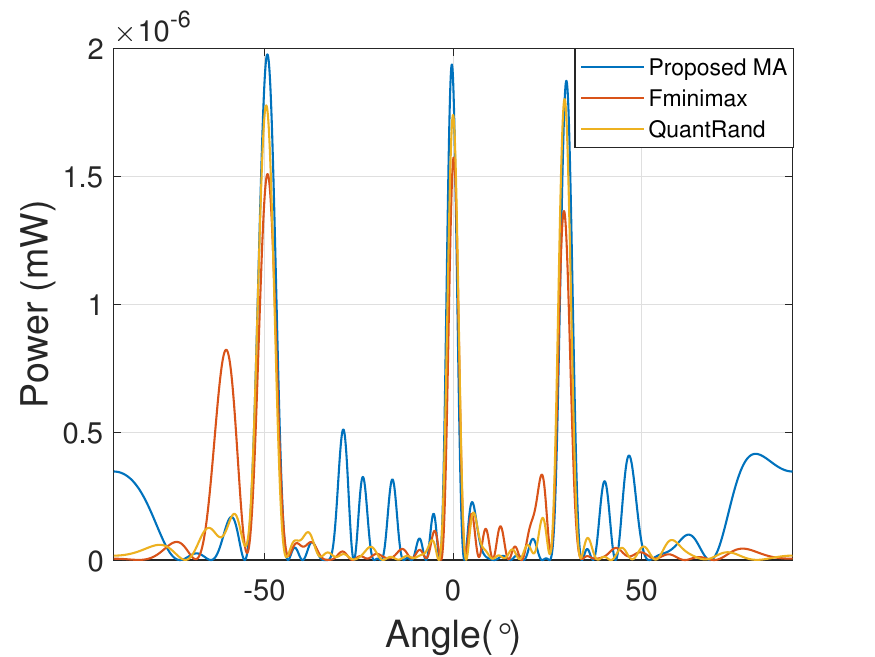}}
  \subfigure[]{
  \label{beam123-3}
  \includegraphics[width=0.75\linewidth]{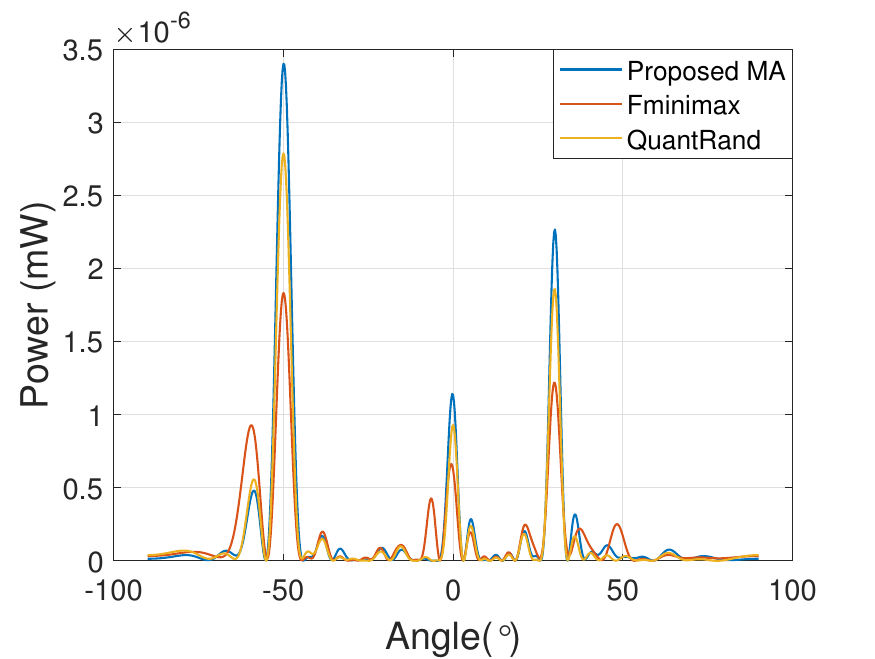}}
  \caption{The comparison regarding the received power with different methods and weight settings. The three UEs are located at ($50^{\circ},180^{\circ}$), ($0^{\circ},0^{\circ}$), ($30^{\circ},0^{\circ}$), respectively. (a) The weights are set as $1,1/2,1/3$. (b) The weights are set as $1,1,1$.}
  \label{beam3}
\end{figure}

\begin{figure}[!htbp]
  \centering
  \subfigure[]{
  \label{powerN}
  \includegraphics[width=0.75\linewidth]{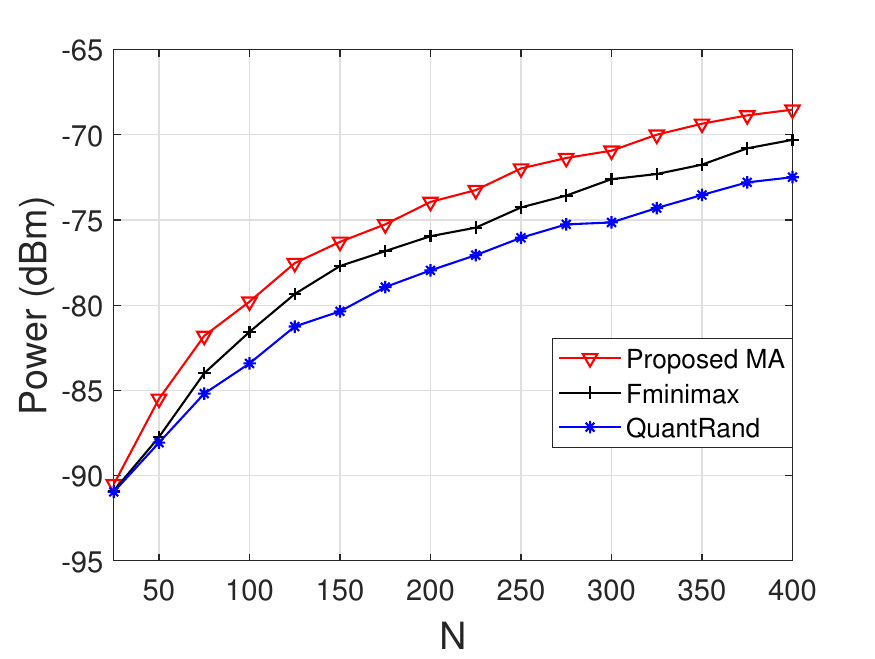}}
  \subfigure[]{
  \label{powerK}
  \includegraphics[width=.75\linewidth]{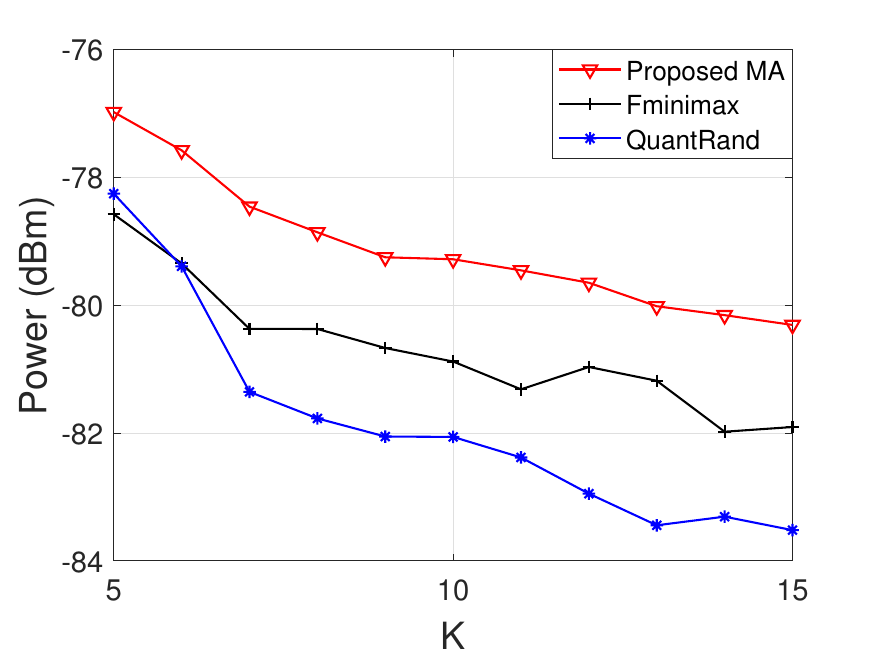}}
  \caption{The minimum received signal power among multiple users employing different algorithms, (a) versus the number of RIS units $N$. (b) versus the number of users $K$.}
  \label{power}
\end{figure}

\begin{figure}[!htbp]
\centering
\includegraphics[width=.75\linewidth]{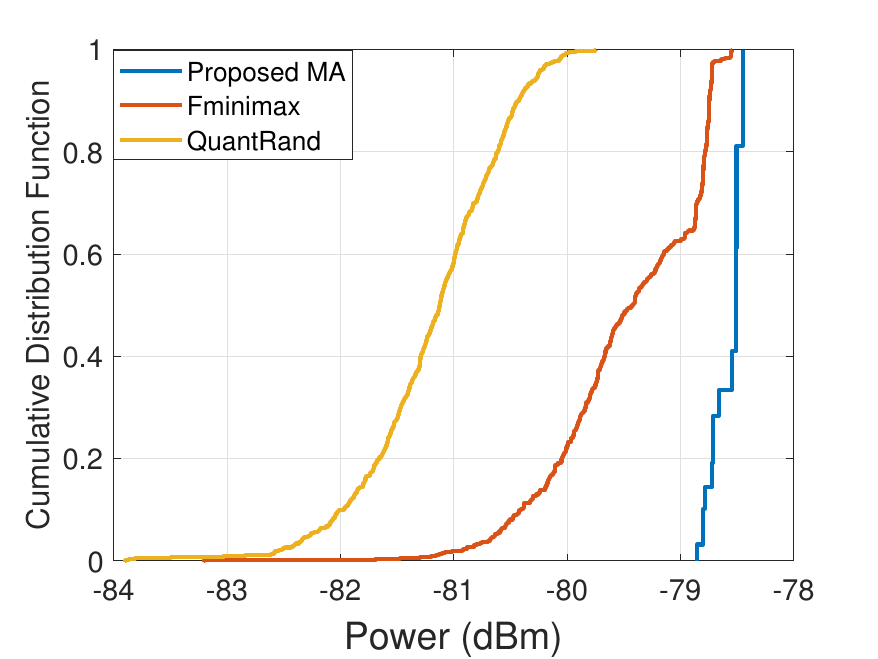}
\caption{Cumulative distribution of received power when $N$ = 100, the number of users is five. }
\label{CDF}
\end{figure}

\subsection{Power Boosting}
We initially provide two examples in Fig.~\ref{beam123-3} to illustrate the radiation patterns corresponding to the MA, Fminimax, and QuantRand algorithms. It is evident that the proposed MA algorithm significantly outperforms the other two at each UE position. Specifically, leveraging a $32\times32$-unit RIS, the proposed algorithm yields an average signal power gain at the three UEs, surpassing the other two algorithms by 11.78\% and 58.90\%, respectively.

For a comprehensive comparison, we systematically record and analyze the received power as the number of units ($N$) increases from 25 to 400 and the number of users ($K$) increases from 5 to 15. To ensure robustness, we conduct 50 trials for each $N$ or $K$ and calculate the average. The results are illustrated in Fig.\ref{power}. In Fig.\ref{powerN}, we present the plots for a fixed $K$ of 10. It can be concluded that as the number of units increases, the received power increases for all algorithms. This makes sense because for an RIS consisting of more units, the induced power is greater. Particularly, these plots underscore substantial power gains achieved by the proposed method. For instance, at $N = 400$, the proposed MA exhibits power gain over 2 dB and 5 dB over Fminimax and QuantRand, respectively.

On the other hand, with a fixed $N =100$, as $K$ increases from 5 to 15, the received power decreases for all algorithms. This is attributed to the fact that, when the number of units is fixed, the induced power remains constant. However, as the number of users to be served increases, the allocated power per user naturally decreases. The plots also demonstrate that the proposed MA algorithm outperforms the competitors.

We finally perform 500 trials with fixed $N=100$ and $K=5$, comparing the performance using the cumulative distribution function. As illustrated in Fig.~\ref{CDF}, the results consistently show the superior performance of the proposed MA algorithm in terms of signal power.

\subsection{Time Complexity}

\begin{figure}[!htbp]
  \centering
  \subfigure[]{
  \label{timeN}
  \includegraphics[width=0.75\linewidth]{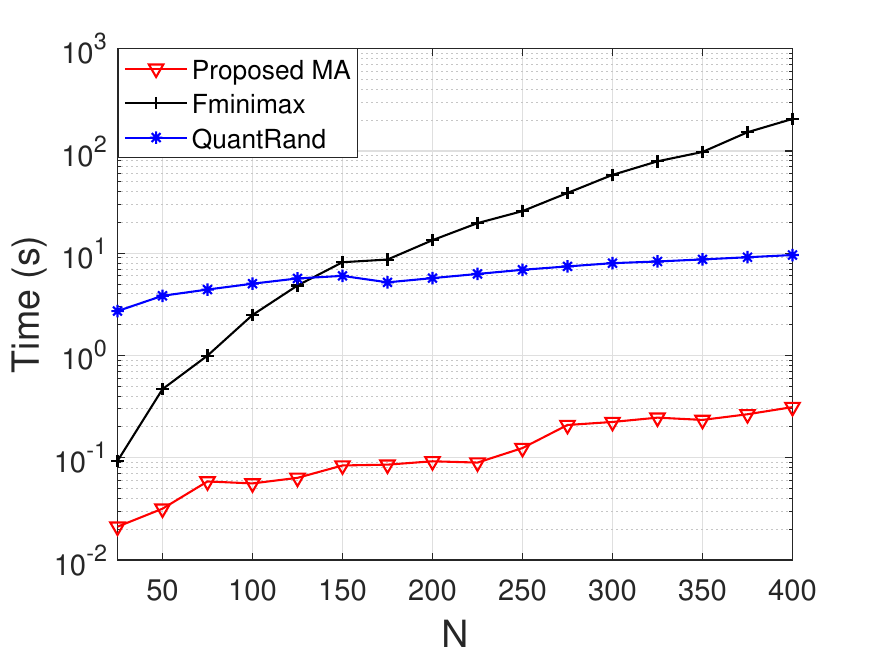}}
  \subfigure[]{
  \label{timeK}
  \includegraphics[width=.75\linewidth]{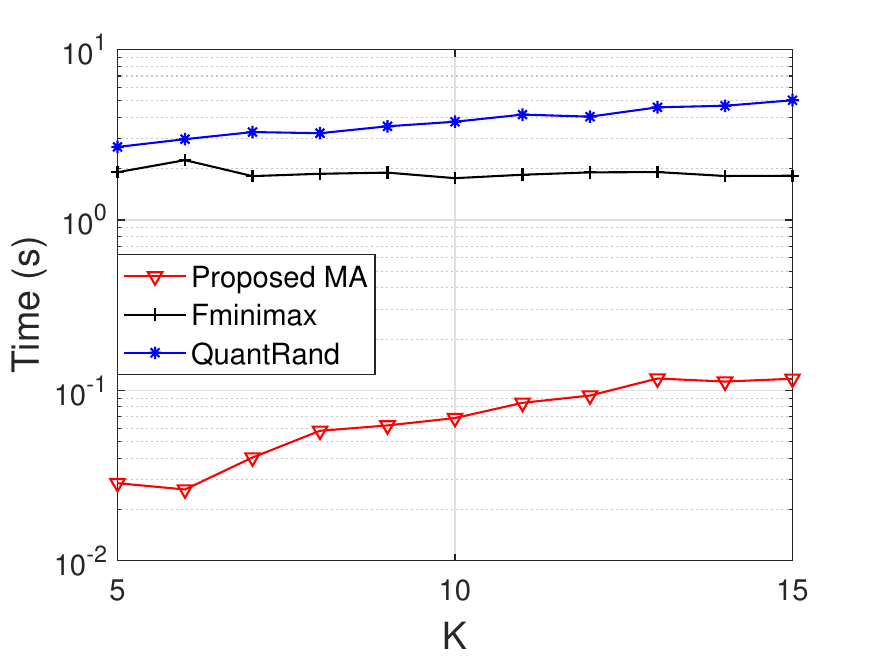}}
  \caption{The proceed time of various algorithms, (a) versus the number of RIS units $N$. (b) versus the number of users $K$.}
  \label{time}
\end{figure}

To evaluate the efficiency of the proposed approach, we conduct tests to assess the execution time as a function of the number of reflecting units ($N$) and users ($K$) in multi-user scenarios. For each value of $N$ or $K$, we perform 50 trials, calculate and plot the average execution time as presented in Fig.~\ref{time}. The findings reveal an increase in the time complexity of all approaches as both $N$ and $K$ increase. Notably, our proposed approach demonstrates the lowest time complexity. These simulations are conducted using MATLAB R2022a on an ${\rm \text{Intel}^{\text R} \text{Core}^{\text TM} }$ i7-12700 CPU at 2.10 GHz with 32 GB RAM.

The results in Fig.~\ref{timeN} demonstrate that the execution time of Fminimax increases substantially as $N$ increases. In fact, for $N = 400$, it costs 205.9 seconds (s) to produce a solution. In contrast, the proposed MA method is much more efficient, when $N = 400$, it takes only 0.31 s on average to find the optimal configuration. The QuantRand algorithm is the second fastest, with an average time of 9 s. 

On the other hand, Fig.~\ref{timeK} illustrates that, with the increase in $K$, the execution time of the QuantRand method significantly rises. Our proposed MA algorithm continues to perform exceptionally well and outperforms the other two algorithms. It obtains the optimal solution in 0.11 seconds even when $K=15$. 
The superior execution-time performance of MA makes it particularly well-suited for large-scale practical implementations.

\subsection{Prototype Experiments}

To assess the effectiveness of the proposed approach in practice, we conduct experiments using a 2-bit RIS-aided communication system. The introduced signal models allow for the direct calculation of phase discrepancies resulting from configurations and interelement path length differences. This characteristic facilitates beamforming without the need for explicit CSI estimation procedures.


\begin{figure*}[!htbp]
\centering
\includegraphics[width=0.9\linewidth]{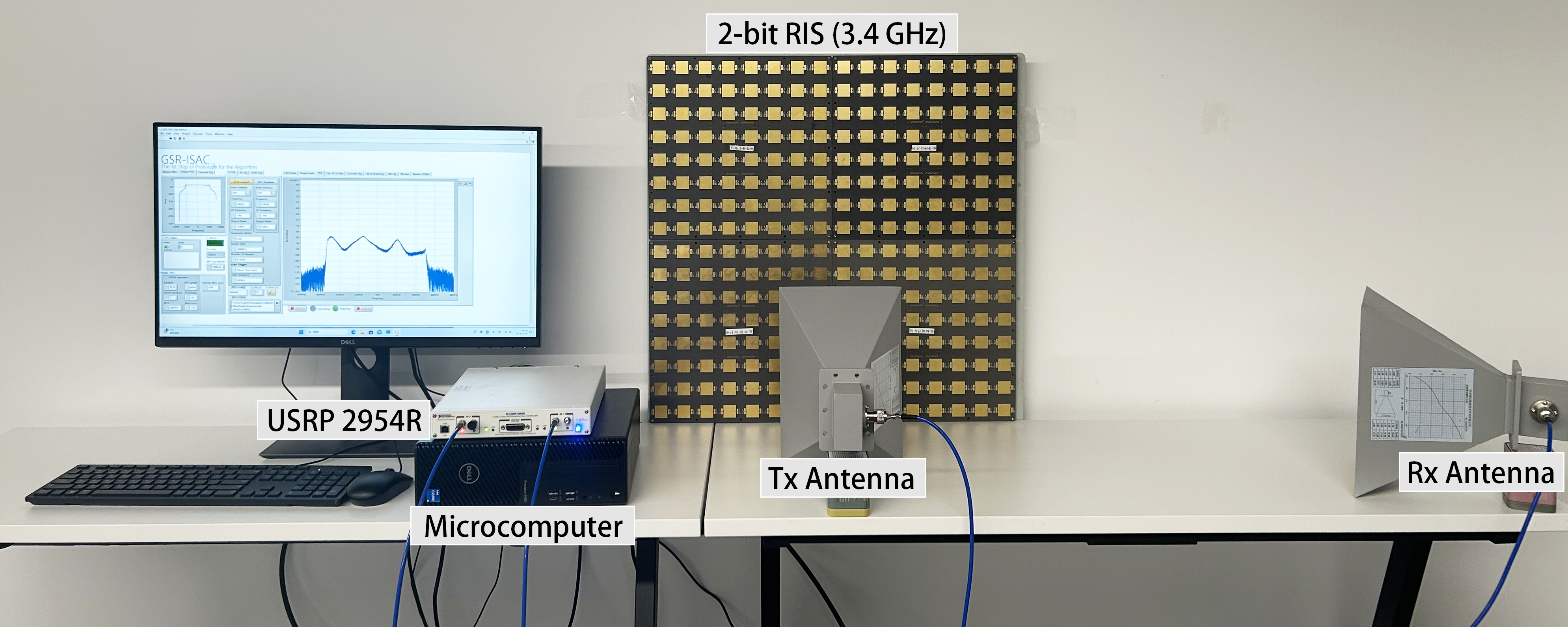}
\caption{The prototype of 2-bit RIS-aided wireless communication system.}
\label{system}
\end{figure*}

The prototype validations are conducted utilizing a 2-bit RIS-aided communication system operating at the central frequency of 3.4 GHz, as illustrated in Fig.~\ref{system}. The RIS is composed of $16 \times 16$ units. A chirp signal with a bandwidth of 100 MHz is transmitted and received by the USRP 2954R using horn antennas. The Tx antenna is positioned at $(0.984 \ \text{m}, 0^{\circ}, 0^{\circ})$. Additionally, three observation points are located at $(6.440 \ \text{m}, 50^{\circ},0^{\circ})$, $(7.925\ \text{m}, 40^{\circ},180^{\circ})$, and $(6.984\ \text{m}, 60^{\circ},180^{\circ})$, respectively.  


\begin{figure}[!htbp]
  \centering
  \subfigure[]{
  \label{57}
  \includegraphics[width=0.65\linewidth]{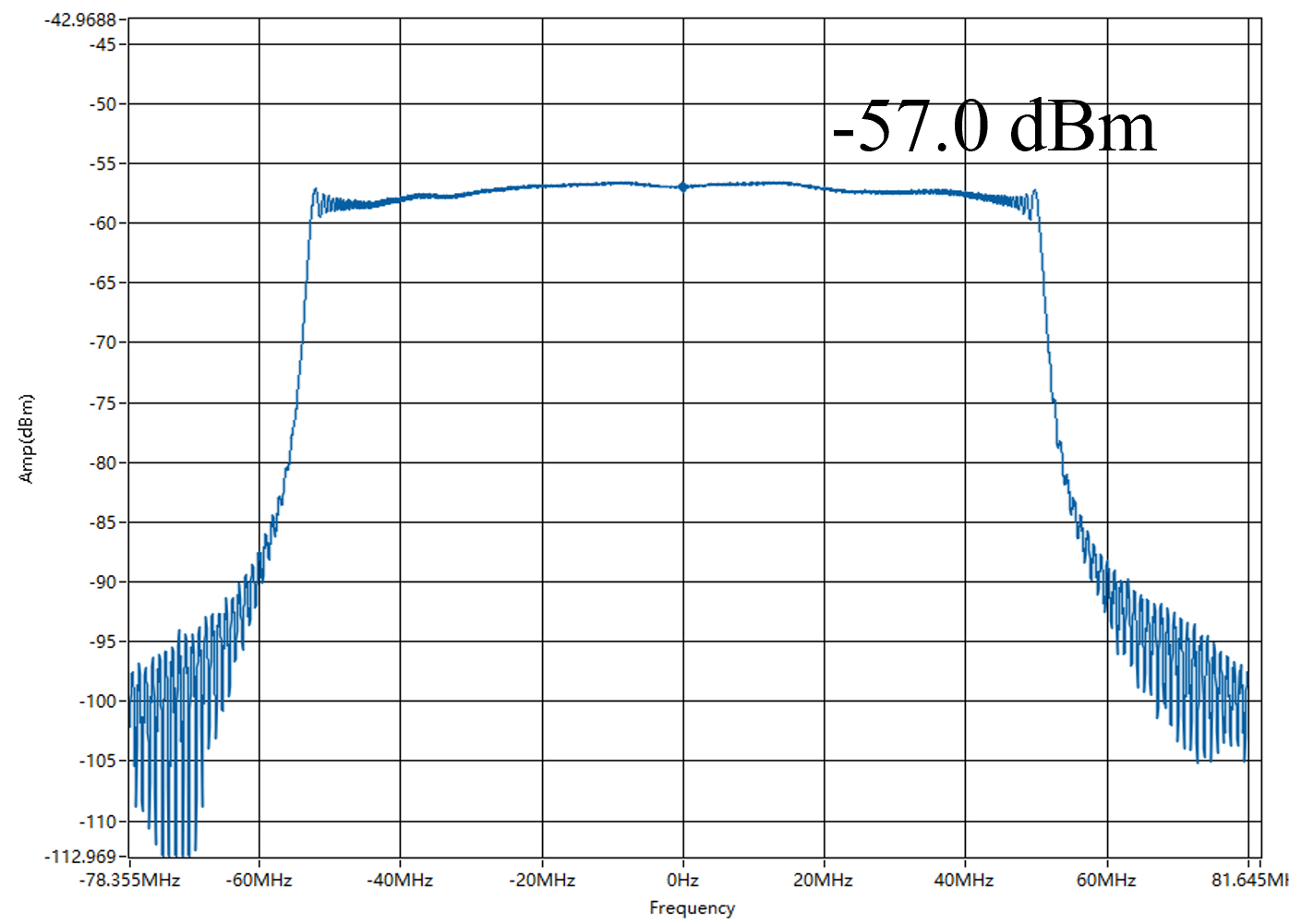}}
  \subfigure[]{
  \label{60}
  \includegraphics[width=.65\linewidth]{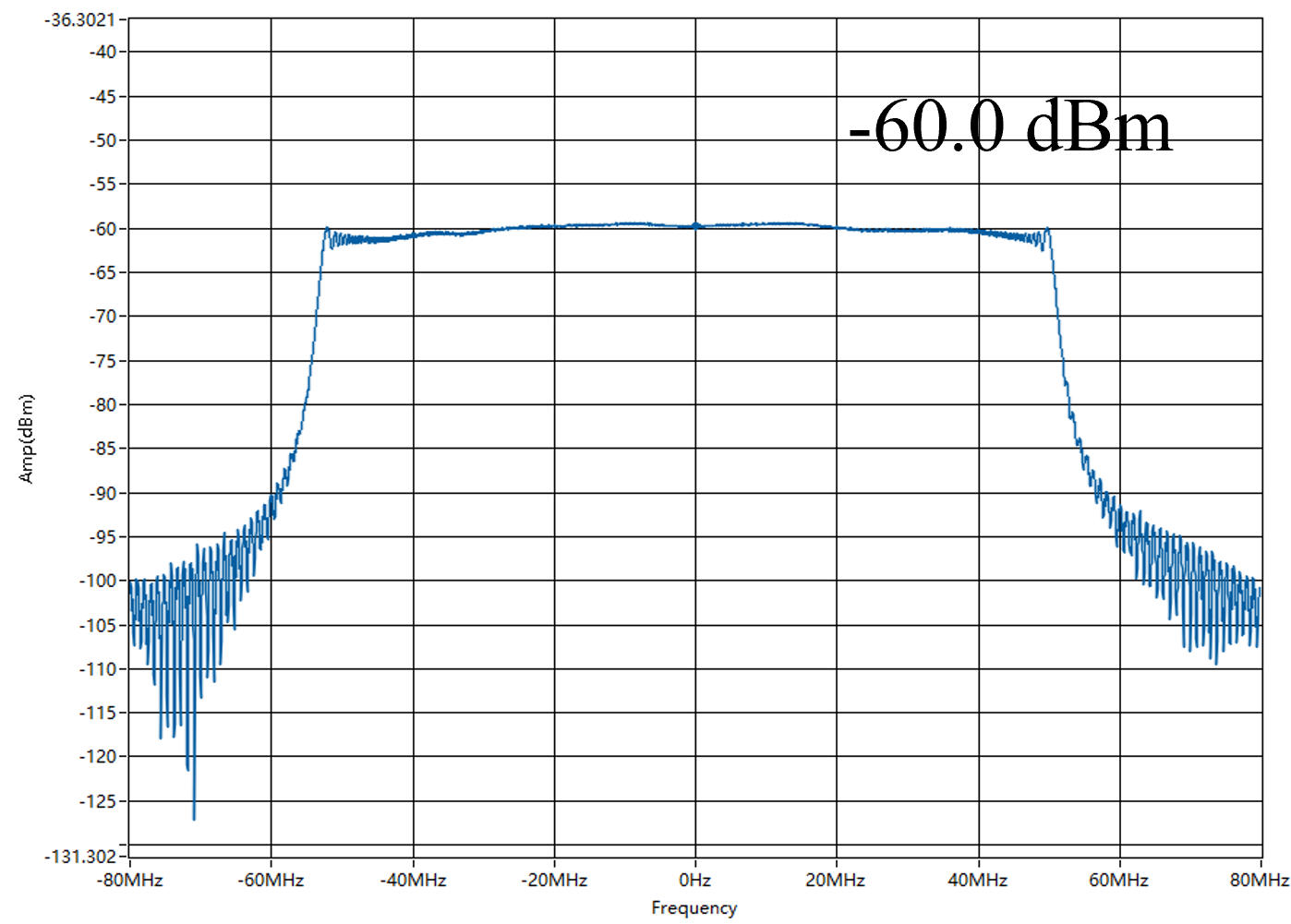}}
  \subfigure[]{
  \label{63}
  \includegraphics[width=.65\linewidth]{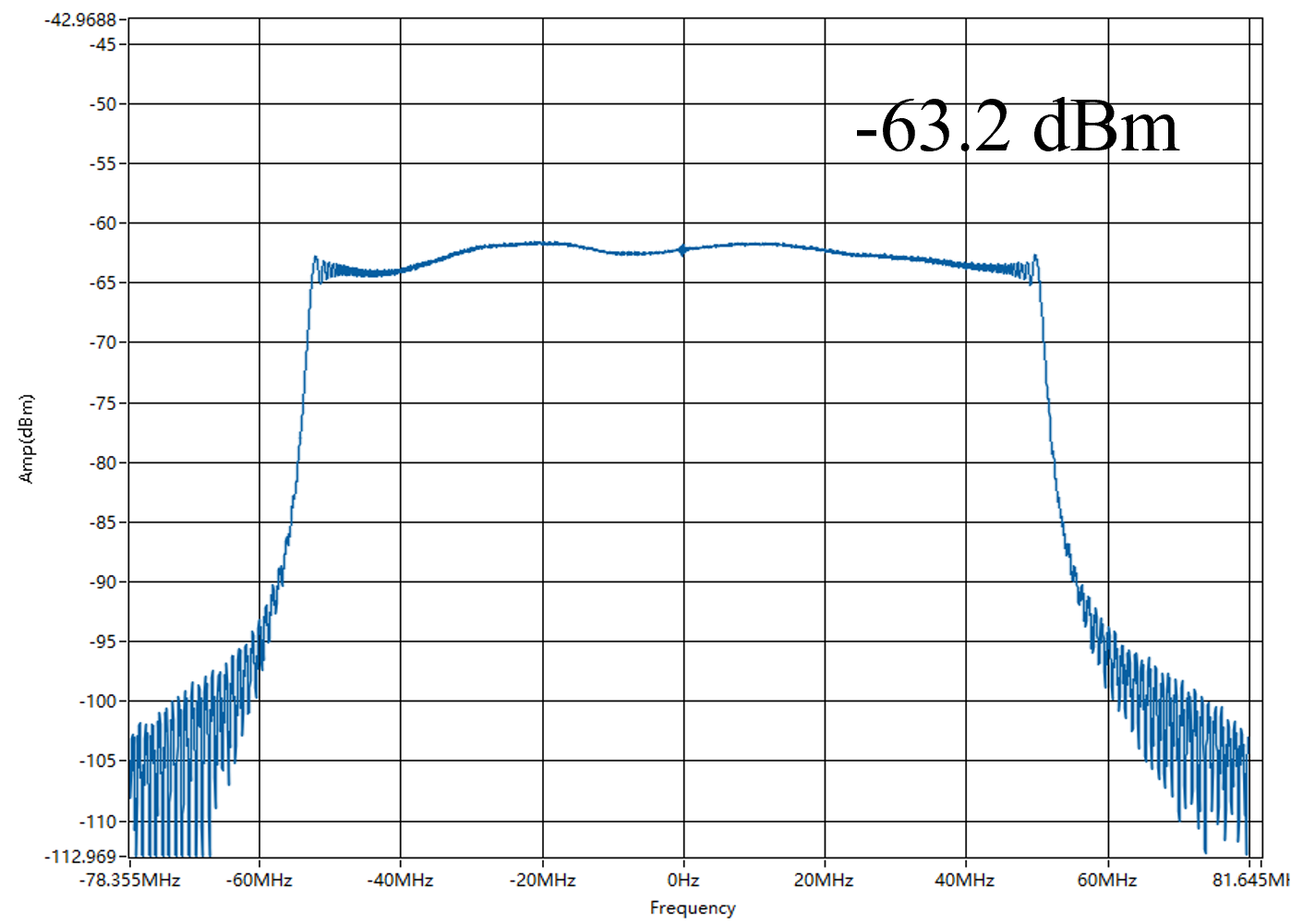}}
  \caption{The actual received power observations with optimized RIS phase under an anticipated weight arrangement of 5, 2, and 1. (a) UE\#1. (b) UE\#2. (c) UE\#3.}    
  \label{experiment}
\end{figure}

%

We initially validate the fair beam allocation functionality of RISs using the proposed MA algorithm to solve the problem (P1). Here, we set $\alpha(r^\text{s}_k) = 1 / (r^\text{s}_k) ^2$ and calculate the steering matrix $\mathbf{H} ( \theta^\text{s}_k )$ appropriately for each $k \in {1, 2, 3}$. Theoretically, the resulting RIS phase configurations ensure equal received power at each observation point. The practical received powers at these points are measured at -59.0 dBm, -59.6 dBm, and -59.2 dBm, respectively. There is a slight deviation, which may be attributed to the discrete nature of the 2-bit RIS phase shifts, as opposed to a continuous spectrum. It can be regarded as equal received power considering an acceptable tolerance. 


Moreover, an additional experiment is conducted to validate the influence of the weight $\alpha(\cdot)$. In this test, we set the weights $\alpha(r^\text{s}_1), \alpha(r^\text{s}_2), \alpha(r^\text{s}_3)$ as $1 / (r^\text{s}_1)^2, 2 / (r^\text{s}_2)^2, 5 / (r^\text{s}_3)^2 $. These weights are chosen to achieve a desired received power ratio of 1:0.5:0.2. After RIS phase optimization by the MA algorithm, the observed received powers at the three users register as -57.0 dBm, -60.0 dBm, and -63.2 dBm, as depicted in Fig.~\ref{experiment}. Similarly, with an acceptable tolerance, the practical power levels are considered in agreement with the desired ratios. 
The prototype experiments validate both the flexible beamforming functionality and the effectiveness of the proposed framework in practice.



\section{Conclusion}\label{Section6}

This paper introduces a comprehensive framework for fair beam allocations through RISs by integrating the Max-min criterion. We introduce realistic models rooted in geometrical optics to characterize input/output behaviors and propose the Moreau-Yosida approximation-based (MA) algorithm for requirements of explicit beamforming functions through RISs. It is worth noting that the framework approach boasts outstanding extensibility, enabling its applicability to a broader class of Max-min optimization problems. Within the proposed beam allocation framework, we explore several fundamental beamforming functions including beam-splitting, fair beam allocation, and wide-beam generation, which had not been thoroughly examined in previous studies. Numerical simulations and prototype experiments substantiate the effectiveness of this work.




%

\bibliographystyle{IEEEtran}
\bibliography{Reference}
%
%
%
%
%
%
%
%
%
%

\newpage

 


\vspace{11pt}


\vfill
\end{document}